\documentclass[12pt,preprint]{aastex}

\slugcomment{Accepted for Publication in ApJ}

\shorttitle{Mid-infrared spectroscopy of submillimeter galaxies}
\shortauthors{Valiante et al.}

\begin{document}

\title{A mid-infrared spectroscopic study of submillimeter
galaxies: luminous starbursts at high redshift}


\author{E. Valiante\altaffilmark{1}, D. Lutz\altaffilmark{1},
E. Sturm\altaffilmark{1}, R. Genzel\altaffilmark{1}, L.J. Tacconi\altaffilmark{1}, M.D. Lehnert\altaffilmark{1},
A.J. Baker\altaffilmark{2}}
\altaffiltext{1}{Max-Planck-Institut f\"ur extraterrestrische Physik,
Postfach 1312, 85741 Garching, Germany \email{valiante@mpe.mpg.de, lutz@mpe.mpg.de, sturm@mpe.mpg.de, genzel@mpe.mpg.de, linda@mpe.mpg.de, mlehnert@mpe.mpg.de} }
\altaffiltext{2}{Department of Physics and Astronomy, Rutgers, The State University of New Jersey, 136 Frelinghuysen Road, Piscataway, NJ 08854-8019 \email{ajbaker@physics.rutgers.edu}}

\begin{abstract}
We present rest frame mid-infrared spectroscopy of a sample of
13 submillimeter galaxies, obtained using the Infrared Spectrograph (IRS)
on board the \textit{Spitzer Space Telescope}. The sample includes
exclusively bright objects from blank fields and cluster lens assisted
surveys that have accurate interferometric positions. We find that the
majority of spectra are well fitted by a starburst template or by the
superposition of PAH emission features and a weak mid-infrared continuum,
the latter a tracer of Active Galactic Nuclei (including Compton-thick ones).
We obtain mid-infrared spectroscopic redshifts for all nine sources detected
with IRS. For three of them the redshifts were previously unknown. The
median value of the redshift distribution is $z\sim 2.8$ if we assume
that the four IRS non-detections are at high redshift. The median for the
IRS detections alone is $z\sim 2.7$. Placing the IRS non-detections at similar
redshift would require rest frame mid-IR obscuration larger than is seen in local
ULIRGs. The rest frame mid-infrared spectra and mid- to far-infrared
spectral energy distributions are consistent with those of local
ultraluminous infrared galaxies, but scaled-up further in luminosity.
The mid-infrared spectra support the scenario that submillimeter
galaxies are sites of extreme star formation, rather than X-ray-obscured
AGN, and represent a critical phase in the formation of massive galaxies.
\end{abstract}

\keywords{galaxies: starburst, galaxies: active, galaxies:
distances and redshifts, infrared: galaxies}

\section{Introduction}
Less than a decade ago, observations with the Submillimeter Common User
Bolometer Array (SCUBA; \citealt{holland99}) on the James Clerk Maxwell
Telescope (JCMT) identified a new and unexpected population of submillimeter 
galaxies (SMGs: \citealt{smail97,barger98,hughes98}). Subsequent surveys with SCUBA and the Max-Planck Millimeter Bolometer (MAMBO; \citealt{kreysa98}) array at the IRAM 30m telescope resolved a significant fraction of the cosmic
submillimeter background into individual sources (see \citealt{blain02},
and references therein).

A detailed understanding of this population has emerged only slowly, due 
to their faintness at all short wavelengths and the difficulty of 
counterpart identification. Photometric  estimates of median redshifts 
around $2.5-3$ \citep{carilli00} are consistent with the recent
determination of a median redshift of $z \sim 2.2$ for the $<50\%$ of
the population accessible to optical spectroscopy \citep{chapman05}. 
The optical redshifts of the radio/optical bright sub-class have been
confirmed in $\sim 15$ cases through CO line detections
\citep{frayer98,frayer99,neri03,greve05,tacconi06}. Despite all these 
efforts, the characterization of the redshift distribution remains 
incomplete, because of the large positional uncertainties for SMGs without 
interferometric counterparts and the large uncertainties in purely 
photometric redshift estimates.

The tendency of SMGs to be faint in X-rays suggests that their
large luminosities result primarily from high star formation rates 
\citep{alexander03,alexander05b}. Moreover, dynamical and gas phase 
metallicity studies indicate that they have high 
(approaching $10^{11} M_{\bigodot}$) baryonic masses
\citep{genzel03,neri03,tecza04,swinbank04,greve05,tacconi06}. 
These high star formation rates and baryonic masses place SMGs 
at the assembly phase of massive galaxies at $z=2-3$, in rapid and
efficient starbursts likely fed by mergers.

The contribution of AGN to the energy output of SMGs has strong 
implications for the origin of the cosmic submm background and the
origin of the correlation between black hole mass and spheroid mass/velocity 
dispersion in local galaxies \citep{ferrarese00,gebhardt00}. Studies of 
the relationship between stellar and black hole mass in submillimeter 
galaxies, under the assumption of accretion at the Eddington rate, 
are consistent with a model where the supermassive black holes in 
SMGs undergo rapid growth to reach the local $M_{\star}-M_{BH}$
relation \citep{borys05}.

SMGs thus mark a pivotal point in the evolution of massive galaxies and their
central black holes. Observational characterisation of their redshift 
distribution, space densities, masses, metallicities, AGN content, and 
structure are all needed to understand their position in the 
hierarchical growth of 
structure and the growth of massive galaxies by accretion of gas and merging of
smaller galaxies. Models of hierarchical galaxy formation  
\citep[e.g.][]{kauffmann99,baugh03} are in the process of adapting to the new
observational constraints including properties of the SMG population 
\citep[e.g.][]{baugh05,bower06}. Extreme objects, such as the SMGs, can 
trace the formation of the $10^{11}M_{\bigodot}$ galaxies already fully
assembled at redshifts of $z=1.6-1.9$ \citep{cimatti04}. SMGs may also
meet the constraint on rapid formation of low redshift massive ellipticals 
inferred from measurements of the $\alpha$/Fe element abundance ratios
\citep{thomas05}. 

Even at low resolution, rest-frame mid-infrared
spectra of galaxies can discriminate between star formation and accretion 
\citep[e.g.][and references therein]{genzel00} on the basis of four distinct 
spectral components observed in dusty galaxies in the local universe
\citep{genzel98,laurent00,tran01}: (i) strong emission
from the polycyclic aromatic hydrocarbon (PAH) features found over a 
very wide range of star forming galaxies; (ii) a variable but usually 
small contribution of an
HII region continuum steeply rising in the $6-15\,\mu{\rm m}$ rest wavelength range;
(iii) a flatter PAH-free AGN continuum, sometimes accompanied by additional $\lambda\geq 10\,\mu{\rm m}$ AGN related silicate emission \citep{siebenmorgen05,hao05}; (iv) absorption features in the $6-8\,\mu{\rm m}$ range as well as the $9.7\,\mu{\rm m}$ silicate absorption feature \citep{spoon04}.

The anticorrelation between PAH feature strength relative to the AGN
continuum and the ionization state of the ionized gas \citep{genzel98,dale06}
strongly supports these low resolution diagnostics.
Using the \textit{Spitzer Space Telescope}, the diagnostics based on the
mid-infrared spectral components of dusty luminous and ultraluminous
galaxies in the local universe can now be applied to dusty SMGs at 
high redshift.  Our understanding of SMGs can be improved in three ways:

{\bf Verification or determination of redshift}. Mid-infrared spectral
features, in particular the narrow aromatic PAH features, if present, 
allow reasonably accurate redshift measurements ($\Delta z \lesssim 0.1$)
even for targets that are very faint at optical wavelengths. They can verify
optical redshifts in cases where these are uncertain due to the faintness of the
optical counterpart, due to the presence of 
multiple candidate counterparts (e.g. 8 of 73 SMGs in the redshift 
study of \citet{chapman05} have multiple radio/optical counterparts), or due to uncertain optical line identifications.
Mid-infrared emission shares with CO line emission a reduced
risk of an erroneous redshift assignment, because these tracers are 
more closely linked than the rest frame UV to the rest frame 
submm/far-infrared emission that dominates the SMG's bolometric
luminosity. They are thus less likely to measure the redshift of a 
misidentified neighbouring source. The difficulties of identification 
with an optical/near-infrared source can be seen for example in the
case of HDF850.1 \citep{downes99,dunlop04}. For sources without known
optical redshifts, new mid-infrared
spectroscopic redshifts strongly reduce the template ambiguities that 
are always possible for model-dependent photometric redshifts. The
advantage of the mid-infrared spectra over CO is that, with existing
instrumentation, there is a larger fractional bandwidth coverage than 
for mm spectroscopy. The disadvantage is that for a
possible $z\gtrsim 4$ tail of the SMG redshift distribution, the main PAH
features leave the wavelength range of sensitive \textit{Spitzer} spectroscopy.

{\bf Evaluation of the relative importance of AGN and star formation} 
using the relative strength of PAH and continuum. Submillimeter sources 
are likely starburst-dominated but more luminous than the threshold above
which most local infrared galaxies are AGN dominated 
\citep{rigopoulou99,veilleux99,tran01}. Furthermore, although massive 
ellipticals in formation should also be forming massive black holes in 
order to produce the $M_{\mbox{\scriptsize{BH}}}-\sigma$ relation
\citep{tremaine02}, evidence for energetically dominant AGN is 
scarce \citep{alexander03}. Evaluating the ratio of starburst and 
AGN contributions will also allow us to investigate trends
with other quantities. We adopt a ratio of $7.7\,\mu{\rm m}$ PAH feature to local
continuum
of 1 as the border between starburst and AGN dominance in the bolometric
luminosity, following the approach of \citet{genzel98} that is well
matched to the rest wavelength coverage and S/N of our data. The rest-frame
mid-infrared continuum, if
isolated from the non-AGN components of the mid-infrared emission, can
provide constraints on the role of AGN, including highly X-ray absorbed
sources \citep{krabbe01,lutz04}. Mid-IR spectra tracing the re-radiation
of absorbed AGN emission thus also constrain the
presence of Compton-thick AGN that are hard to separate from inactive
objects in current high redshift X-ray data, because their X-ray photon
statistics can be insufficient for clear identification of reflected
AGN emission, or because the AGN may be fully covered.

{\bf Constraining physical conditions in the starburst.} By analogy with 
local galaxies, dilute/cool regions are expected to show a larger PAH 
contribution to the total infrared emission and a cooler rest frame 
far-infrared peak than denser starbursts \citep[e.g.][]{dale02}.
Very compact regions can show absorption features like those of water 
ice and hydrocarbons. Some of the most luminous local ULIRGs exhibit 
strong continua with superposed emission features deviating from a 
canonical PAH feature shape \citep{tran01,spoon04}. Similar features may be found in SMGs.

\textit{Spitzer} spectra are also important to test the popular assumption
\citep[e.g.][]{hughes98} that the overall spectral energy 
distributions (SEDs) of SMGs are similar to those of local star forming 
Ultraluminous Infrared Galaxies (ULIRGs).

In \S 2 we discuss the properties of our sample and in \S 3 observations 
and data analysis. \S4 presents the SMG spectra and discusses the results of 
their spectral decomposition. In \S5 we discuss the implications for the
three main questions raised above.

\section{Sample Selection}
Among the several hundred SMGs detected up to now, our sample comprises
exclusively bright ($S_{850\,\rm{\mu m}} \geq 4.5\,\mathrm{mJy}$) objects from blank field and cluster lens assisted surveys
that have accurate positions from follow-up radio or mm interferometry.
Objects from radio pre-selected surveys are not included because they may
be biased in redshift and AGN content. No constraint on mid-infrared
photometric flux, which could potentially introduce biases on SED and energy source, has been applied to this Cycle 1 sample.

The 13 targets in our sample (Tab.~\ref{observations}) cover the full range of properties of currently known bright SMGs. The ratio of optical/near-infrared and submm fluxes, for example, varies by almost two orders of magnitude (Tab.~\ref{sample}). In our sample there are 6 objects with published optical redshifts \citep{barger99,ledlow02,chapman05}, in particular 3 cases where the redshift is confirmed by CO interferometry \citep{neri03,greve05}, and 7 sources for which no optical redshift is available due to their faintness in the optical/NIR. Including these objects in our study reduces biases that may affect a sample with measured optical redshifts only. The redshift distribution of a sample with optical redshifts will tend to avoid the optical ``spectroscopic desert'' ($1.2\lesssim z\lesssim 1.8$). It may also favour sources with unobscured or mildly obscured AGNs because of their strong rest frame UV line emission. We include sources selected at $850\,\mu{\rm m}$ (SCUBA) as well as sources selected at $1.2\,{\rm mm}$ (MAMBO), to minimize selection effects due to redshift or dust temperature. The sample encompasses the five brightest MAMBO sources with interferometric positions known before the \citet{voss06} observations, and a number of bright SCUBA sources with and without redshifts.

\section{Observations and Data Analysis}
We obtained low resolution ($\lambda /\Delta \lambda \sim $ 60-120\footnote{
$\Delta \lambda$ is approximately constant as a function of $\lambda$}) long
slit spectra using \textit{Spitzer} IRS \citep{houck04} in the staring mode. The detector is a $128\times128$ Si:Sb (LL1 and LL2 modules) or Si:As (SL1 module) Blocked Impurity Band (BIB) array.

The rest wavelength range required to detect mid-infrared PAH features and/or silicate absorption, encompasses at a minimum wavelengths from $5$ to $10\,\mu{\rm m}$. Observations are summarized in Table~\ref{observations}. Most of the sources were observed in the LL1 $19.5-38.0\,\mu{\rm m}$ module for 30 cycles of 120\,s ramp duration. The total on-source integration time was 2 hours from addition of the independent spectra created by the telescope nods along the slit. For objects with $z<2.7$ and most objects with unknown redshifts we added the LL2 $14.0-21.3\,\mu{\rm m}$ range with 15 cycles of 120\,s to ensure the required rest wavelength coverage for all plausible redshifts. In these cases, the slit on-source integration time for LL2 was 1 hour. Because of its low redshift, SMMJ16369$+$4057 was observed in the LL2 range for 10 cycles of 120\,s, as well as in the SL1 $7.4-14.5\,\mu{\rm m}$ module for 5 cycles of 240\,s (2400\,s total on-source integration time). With this observation plan, we could in principle detect both the $6.2\,\mu{\rm m}$ and $7.7\,\mu{\rm m}$ PAH features of a starburst-like spectrum for a redshift range $1.4\lesssim z\lesssim 3.8$. For lower redshifts, we would observe only the $7.7\,\mu{\rm m}$ maximum, which fully exits the observed window at $z \sim 1$, but the longer wavelength 11.3 and $12.7\,\mu{\rm m}$ PAH features might still provide useful information in such a case.

We reduced the data as follows. We subtracted, for each cycle, the two nod
positions of the pipeline 14.0.0 basic calibrated data frames. In the difference
just calculated, we replaced deviant pixels by values representative of
their spectral neighborhoods. We subtracted residual wavelength dependent
background, measured in source-free regions of the two dimensional
difference spectra. In averaging all the cycles of the 2-dimensional
subtracted frames, we excluded values more than three times the local
noise away from the mean. The calibrated 1-dimensional spectra for the
positive and the negative beams were extracted using the optimum extraction
mode of the SPICE analysis package (version 1.4.1), and the two
1-dimensional spectra averaged in order to obtain the final spectrum.

Since our measurements are close to the sensitivity limit of the instrument,
for all data analysis we cut away the long and short wavelength ends of
each module, where the noise is much higher and no signals are detected.

In order to do a formal $\chi^2$ template fit for our sources, it is
essential to determine the uncertainties on each point of the spectra.
We used a processed and deglitched image of one of the undetected sources 
(the result was comparable for all of them). The uncertainty for each 
spectral point was calculated as the noise for each pixel times the square 
root of the number of pixels in the resolution element, which increases with
wavelength. The noise for each pixel has been assumed to be the clipped 
standard deviation on a row of 20 pixels, all at the same wavelength. The 
resolution element is defined by the IRS pipeline, subdividing the region 
of the array covered by the spectrum into a sequence of trapezoid-shaped
elements. This estimate may not fully reproduce the absolute noise 
level of a given observation, due to variations in strength of
zodiacal light with position and epoch, but is a good approximation
of the change of noise as a function of wavelength. 

\section{Results}

\begin{figure}[p!]
\centering

{\includegraphics[width=5.2cm,keepaspectratio]{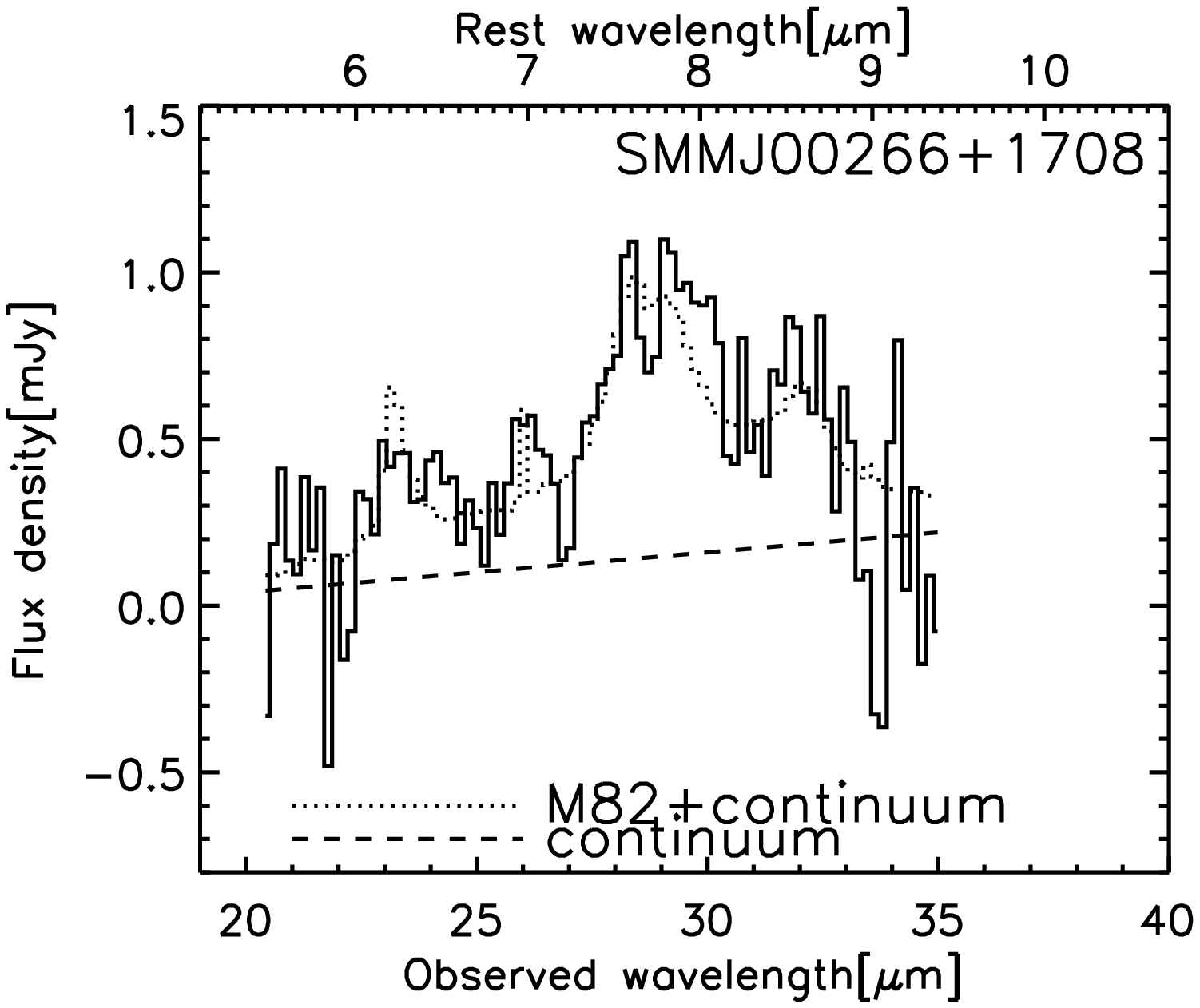}}
{\includegraphics[width=5.2cm,keepaspectratio]{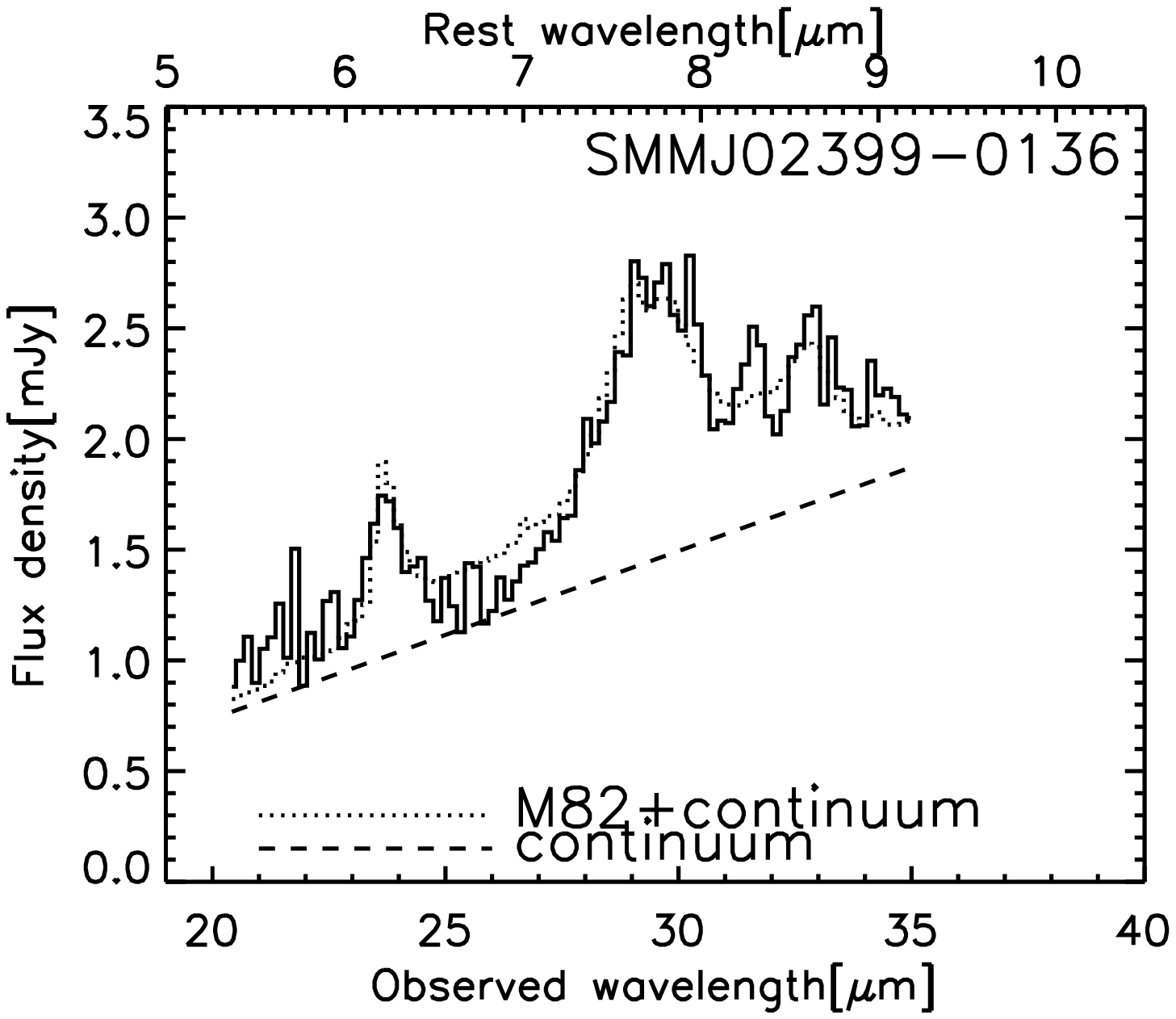}}
{\includegraphics[width=5.2cm,keepaspectratio]{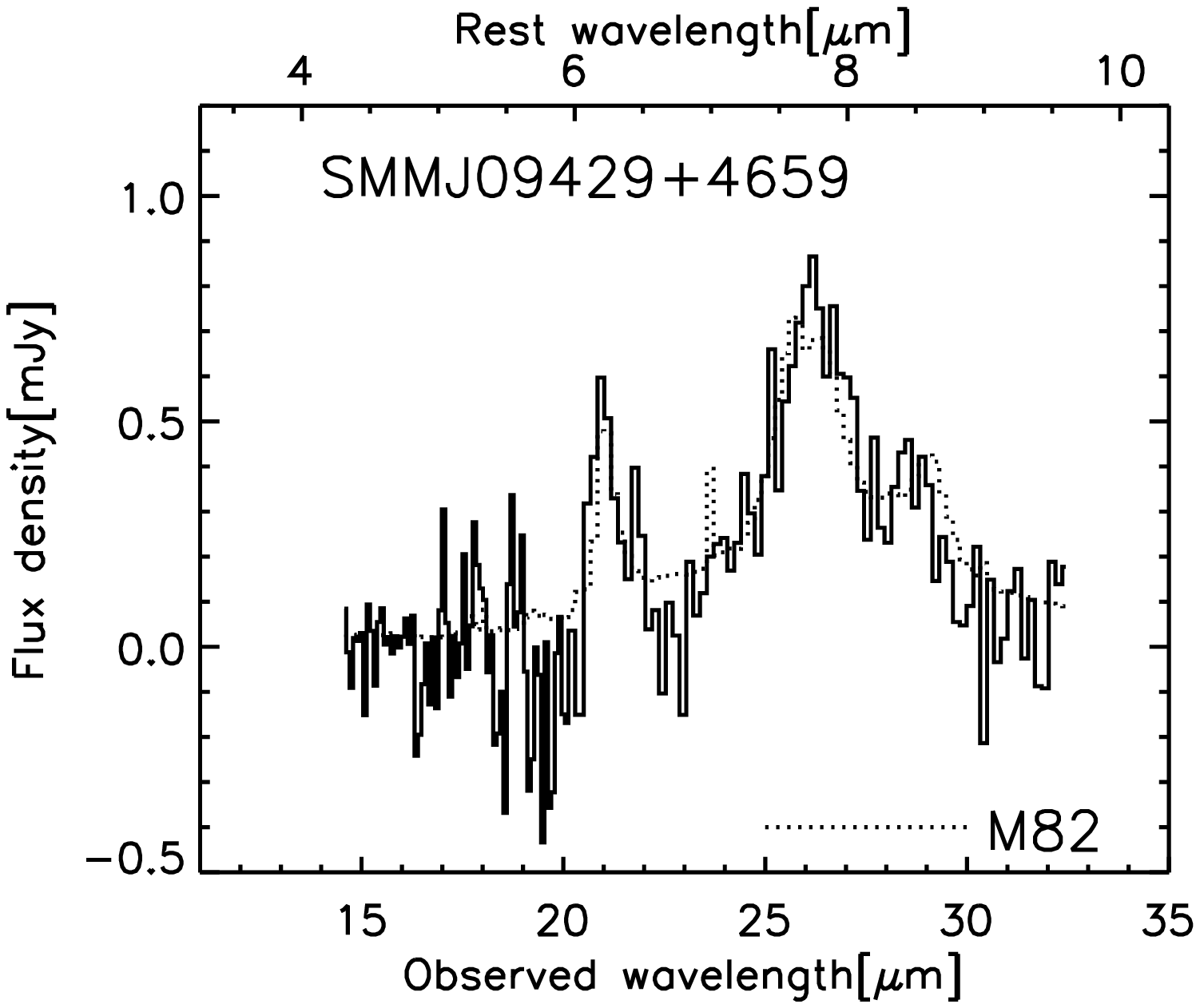}}\\[3mm]
{\includegraphics[width=5.2cm,keepaspectratio]{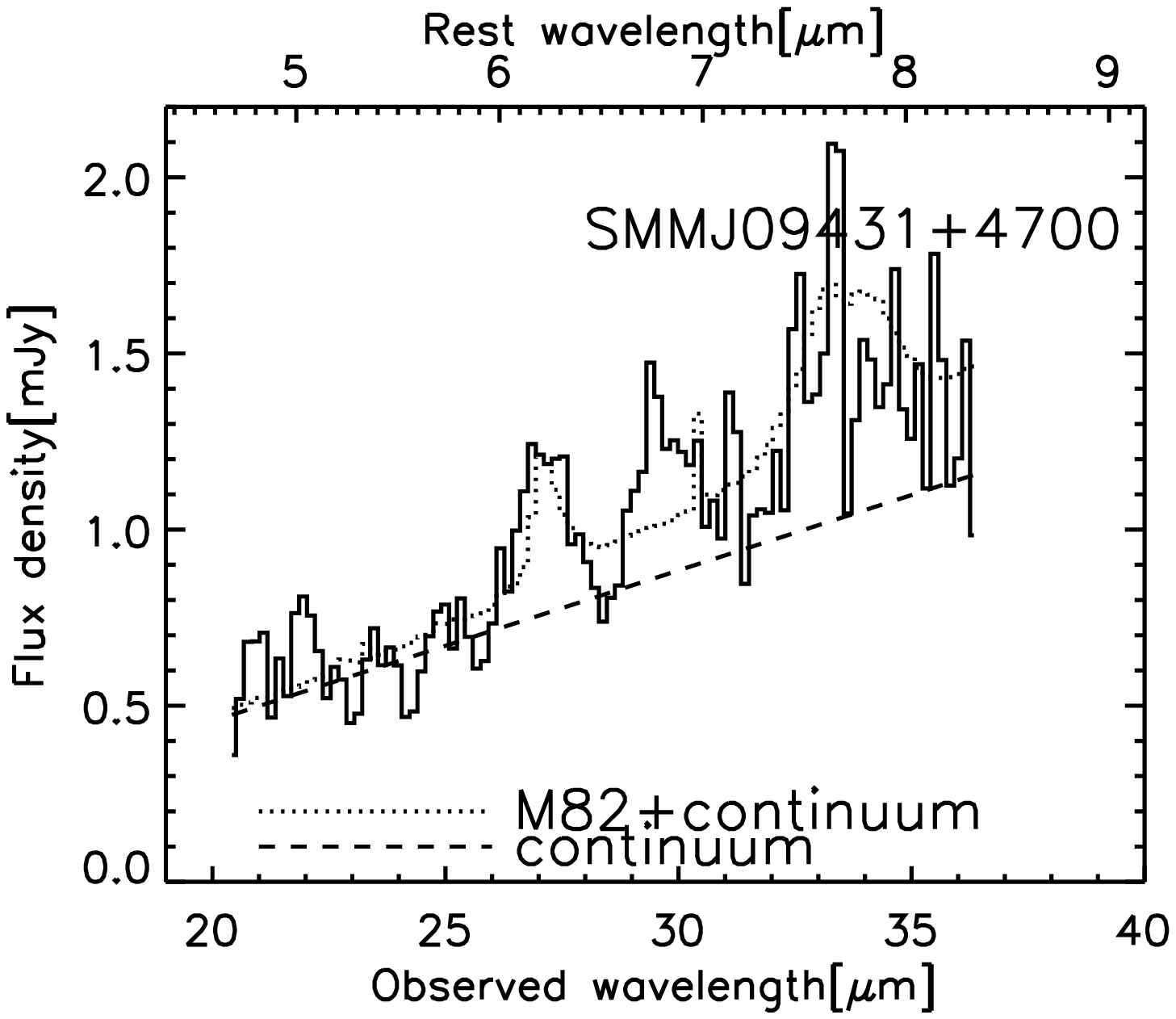}}
{\includegraphics[width=5.2cm,keepaspectratio]{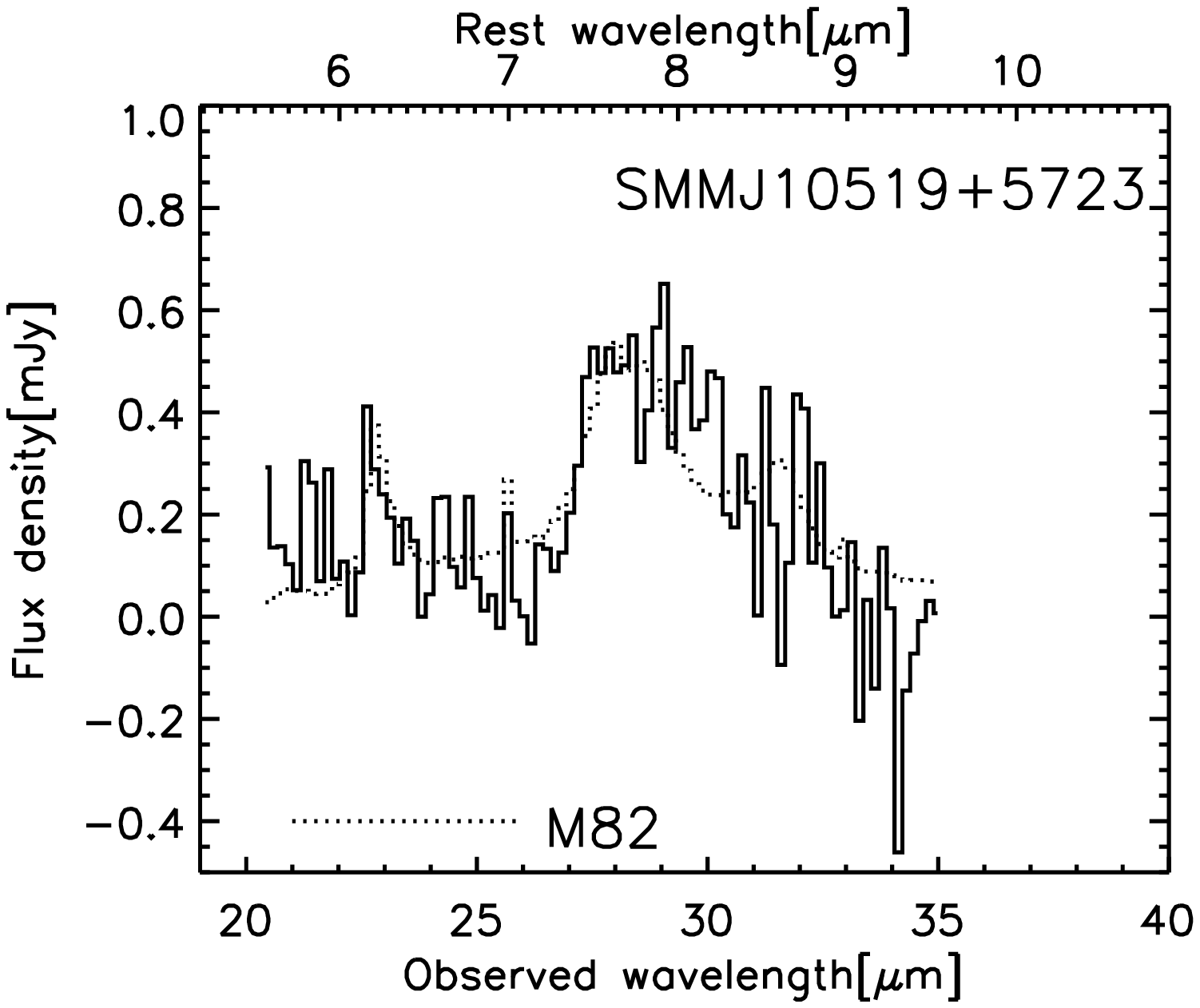}}
{\includegraphics[width=5.2cm,keepaspectratio]{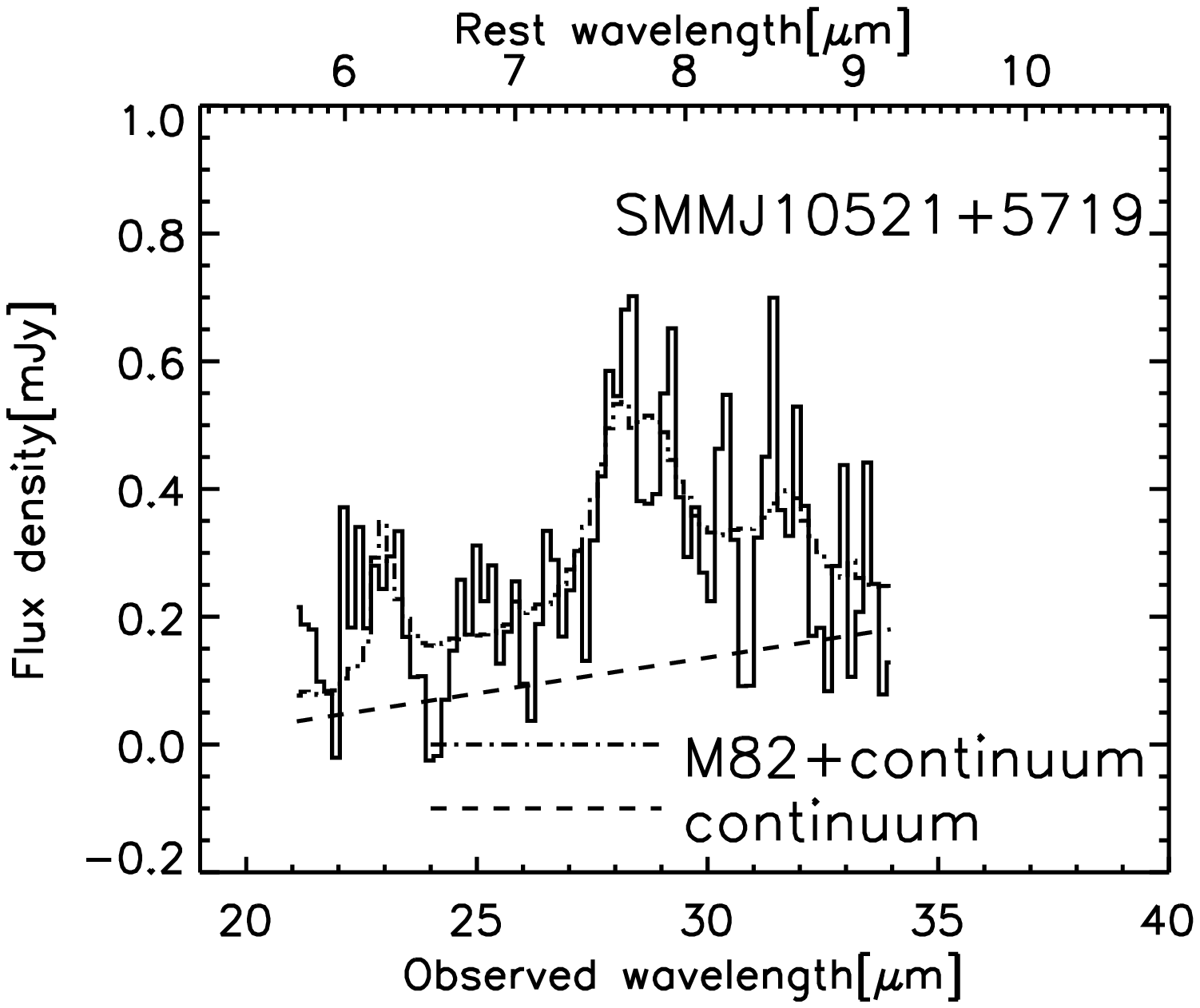}}\\[3mm]
{\includegraphics[width=5.2cm,keepaspectratio]{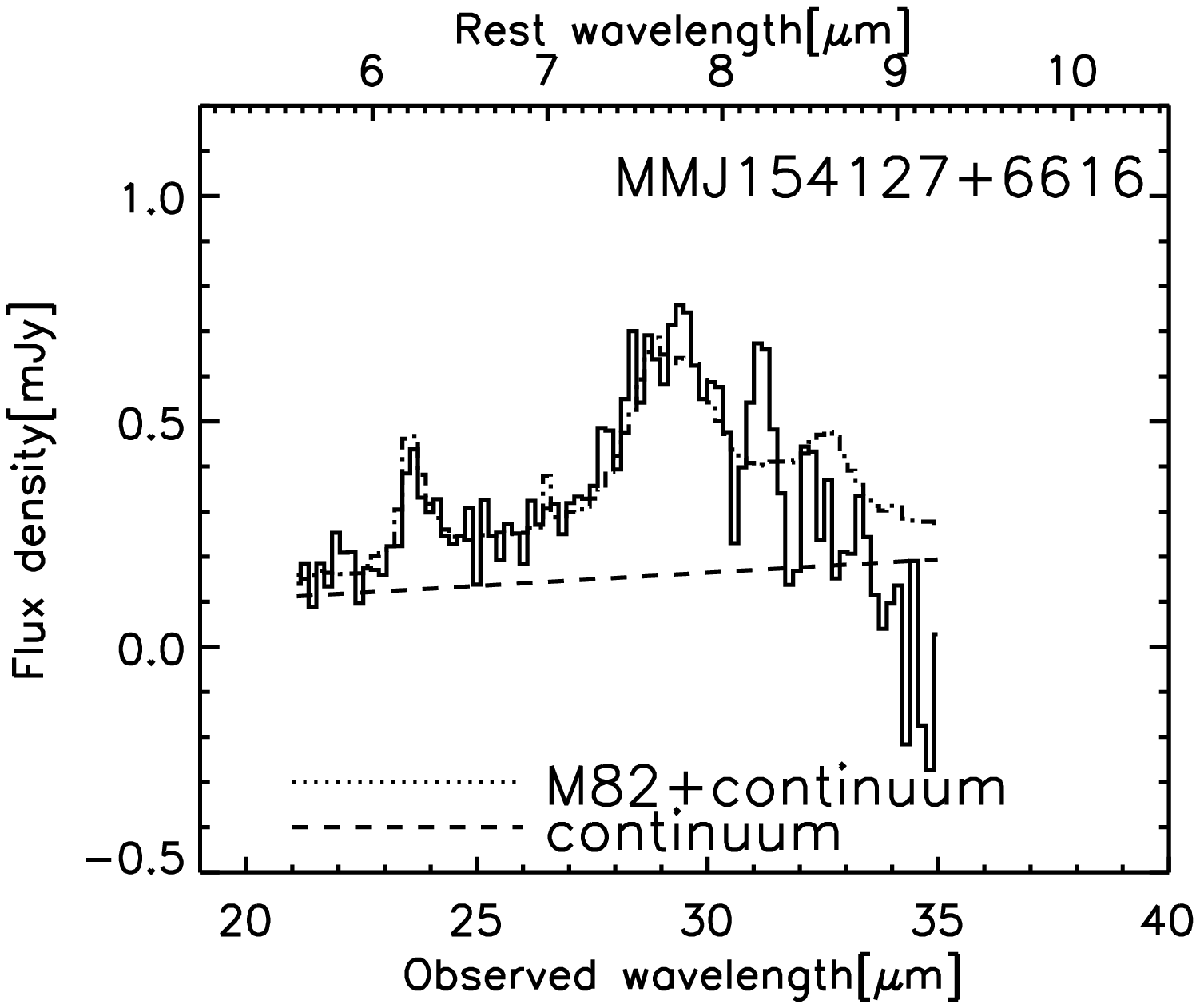}}
{\includegraphics[width=5.2cm,keepaspectratio]{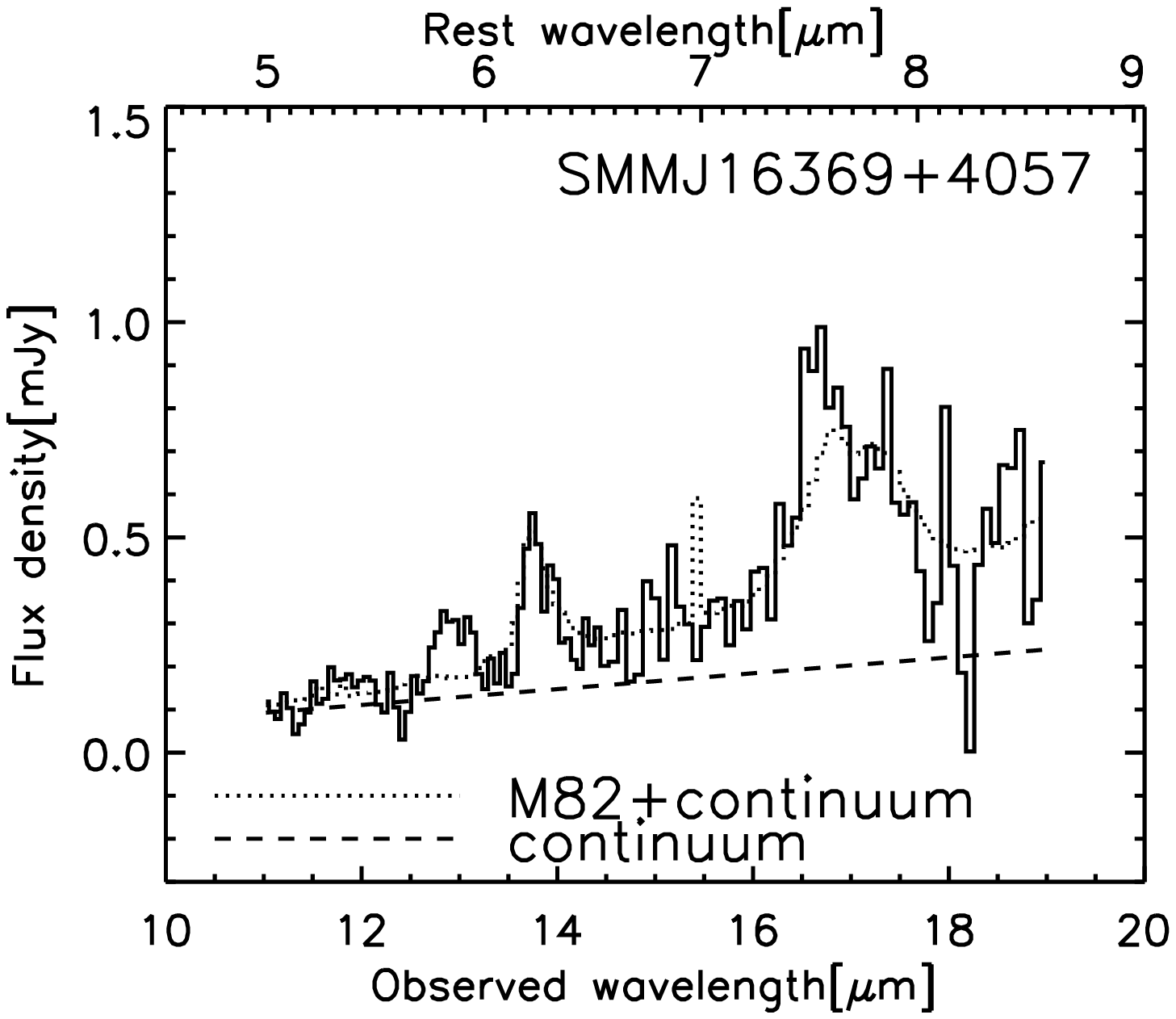}}
{\includegraphics[width=5.2cm,keepaspectratio]{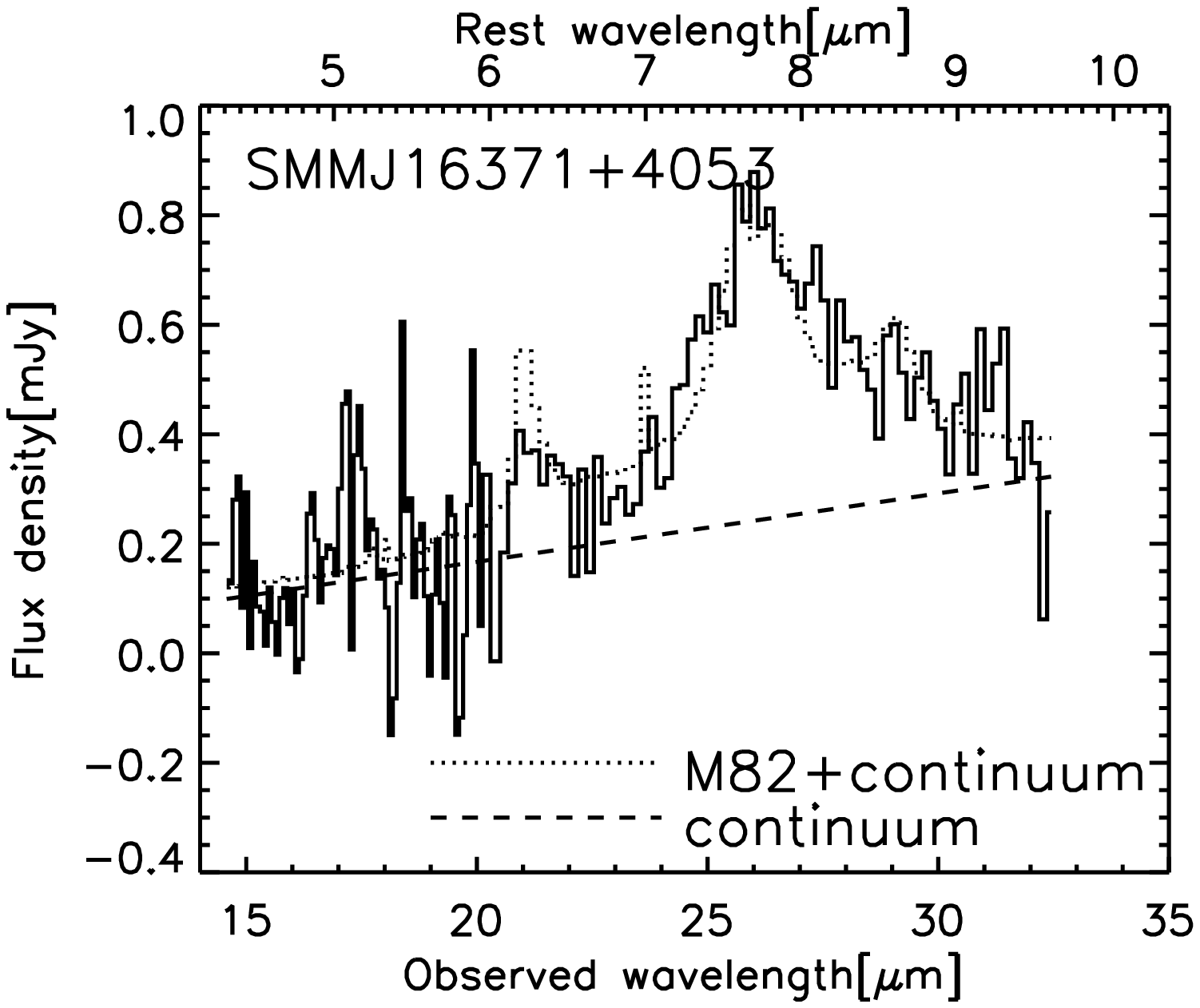}}\\
{\includegraphics[width=4.cm,keepaspectratio]{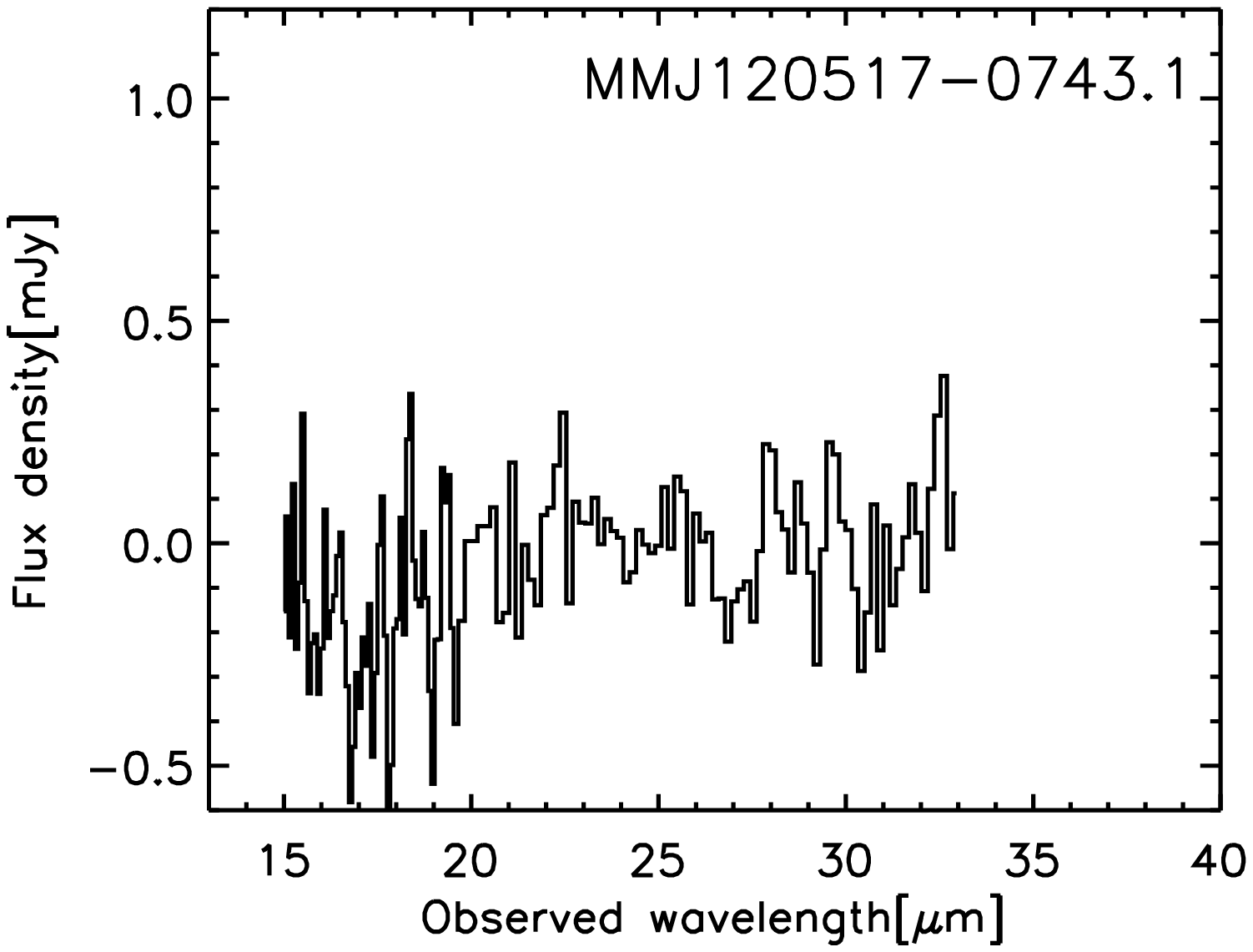}}
{\includegraphics[width=4.cm,keepaspectratio]{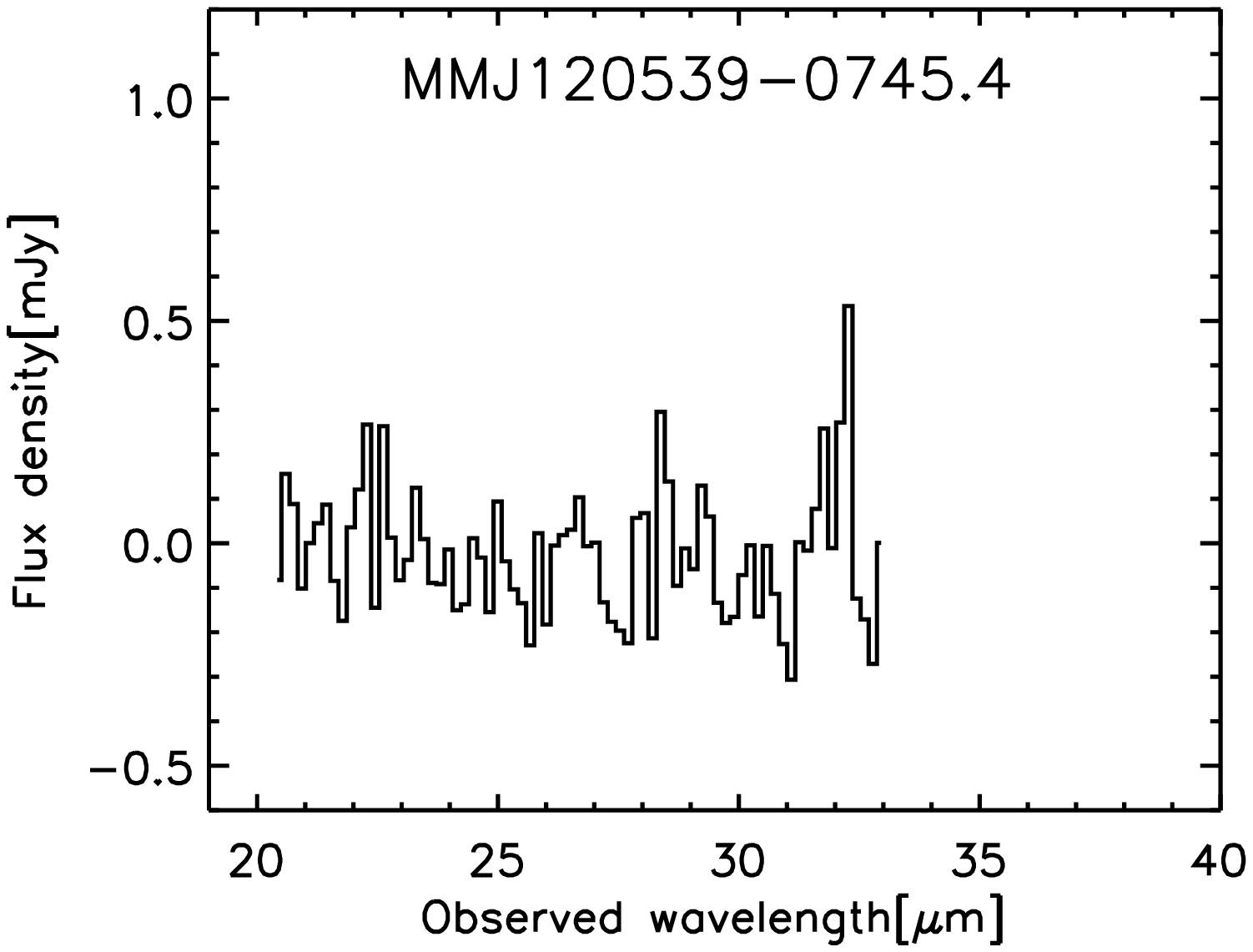}}
{\includegraphics[width=4.cm,keepaspectratio]{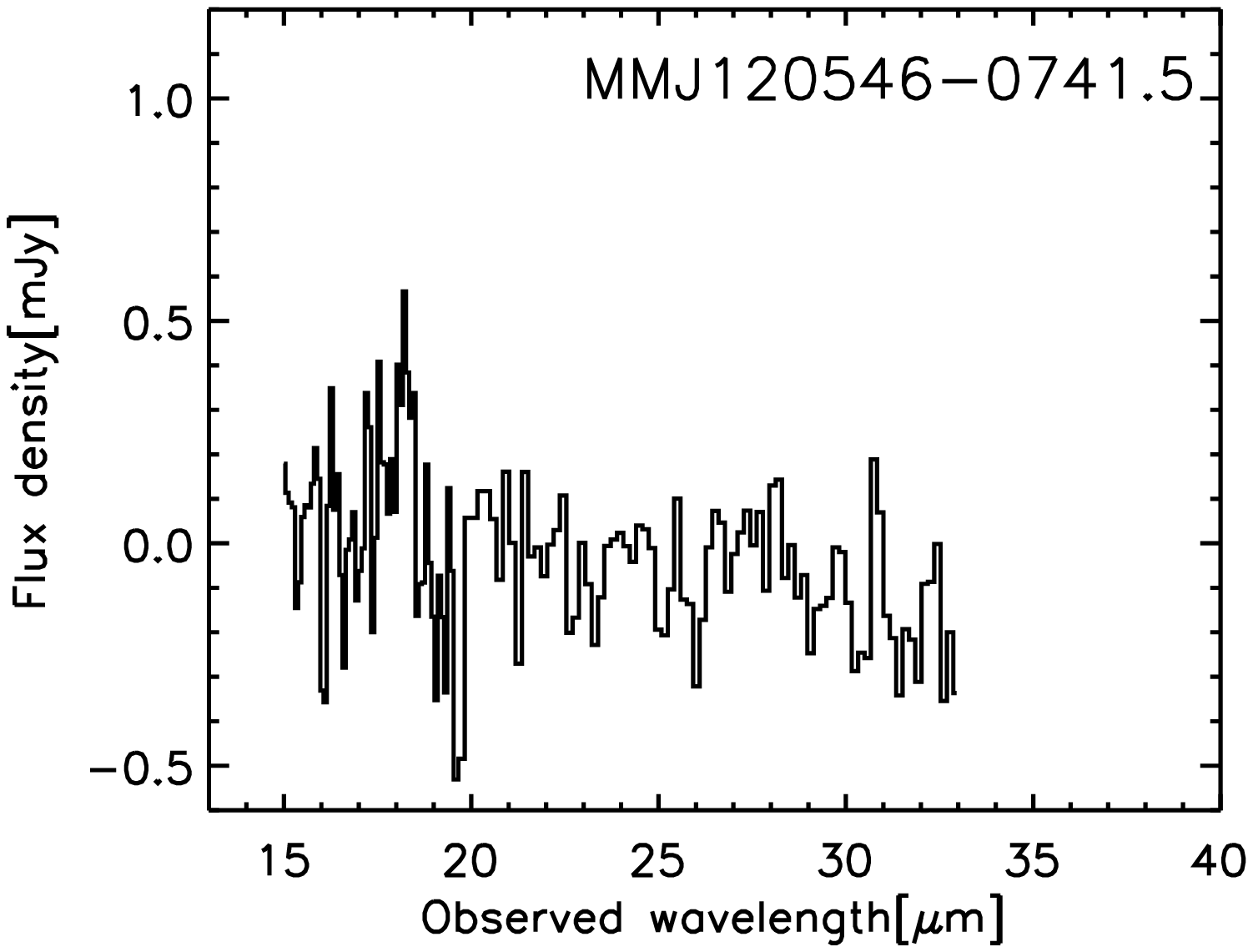}}
{\includegraphics[width=4.cm,keepaspectratio]{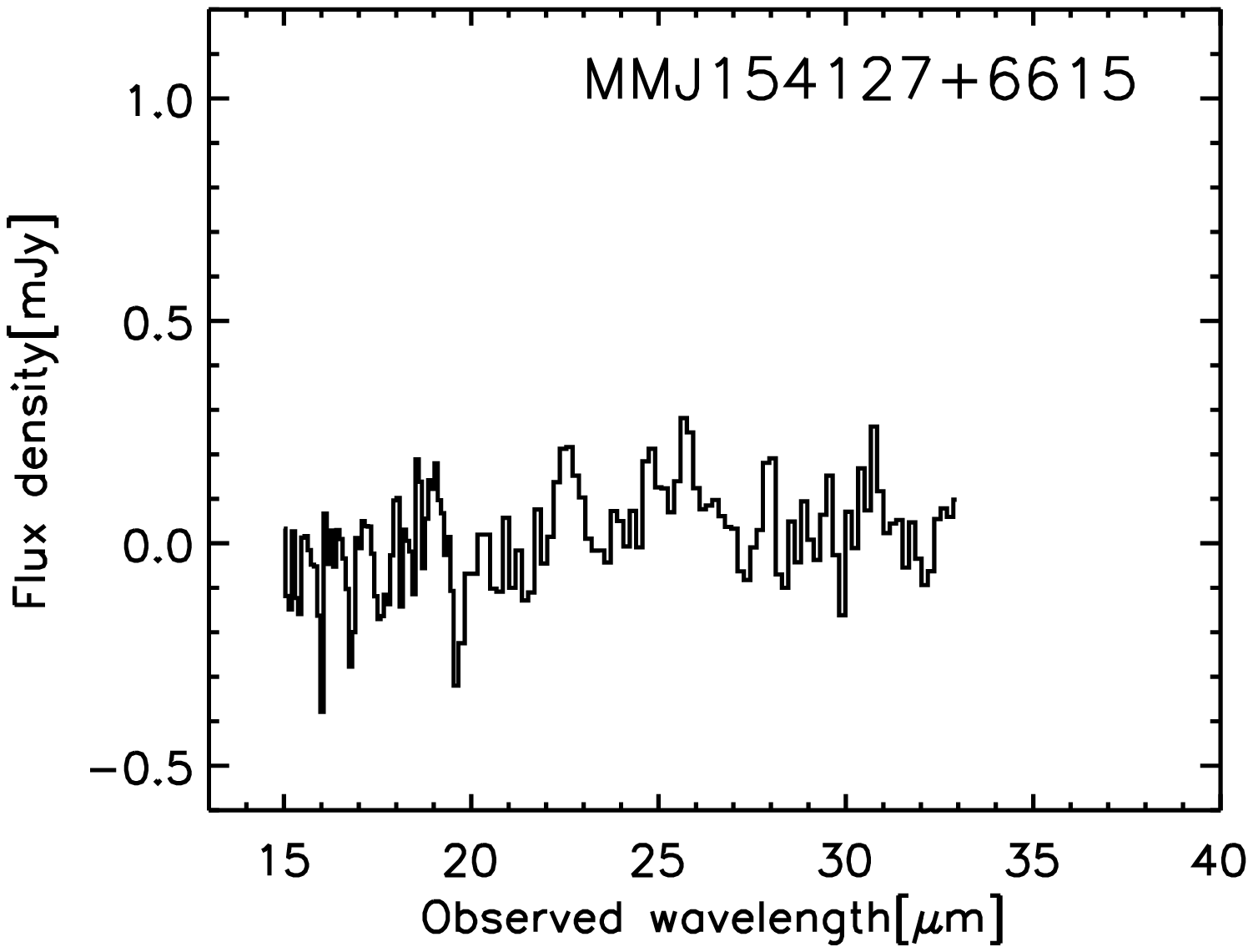}}
\caption{\textit{Spitzer} IRS low resolution spectra (solid lines) of the sample
galaxies. The detected sources are shown together with the best template fit (dotted and dashed line). Their redshifts are listed in Table~\ref{sample}. The last four spectra show the IRS non-detections.}
\label{fit}
\end{figure}

The spectra of the nine detected sources and the four non-detections
are shown in Fig.~\ref{fit}. As quantified by spectral fits below, the
detections can be well reproduced by combinations of PAH features and continua. In one or two cases a heavily absorbed continuum is a possible alternative. We
do not see evidence for strong continua like those seen in local QSOs (which also show silicate emission outside our rest wavelength coverage, see
\citealt{siebenmorgen05,hao05,sturm05}). These individual classifications
are strongly supported by the PAH-dominated average of the SMG spectra
(Fig.~\ref{average}), which shows much larger PAH equivalent widths
than those seen in QSOs \citep{schweitzer06}.

In spectra of faint sources with poor S/N, only spectral features with 
sufficiently large equivalent widths can be used for redshift determination. 
At one extreme, these features are the strong PAH emission features of 
starburst galaxies. These features appear whenever the interstellar 
medium is exposed to moderately intense UV radiation \citep[e.g.][]{draine03}.
The other extreme shows strong absorption features, the strongest being 
silicate absorption. Examples of PAH features in high-redshift galaxies
observed with the IRS are in \citet{teplitz05}, \citet{yan05},
\citet{lutz05} and \citet{desai06}. An extremely absorbed local source is 
IRAS F00183$-$7111 \citep{tran01}, whose IRS spectrum is shown in
\citet{spoon04}. Similar spectra have been observed in optically obscured 
high redshift $24\,\mu{\rm m}$ sources \citep{houck05, weedman06}.

Redshifts can be determined by using either the set of strong PAH emission 
features or, for absorbed spectra, the $8\,\mu{\rm m}$ maximum and nearby silicate
absorption. The strongest PAH feature is at $7.7\,\mu{\rm m}$ (rest frame), so a
similar redshift ($\Delta z \sim \pm 0.1$) would be derived even if it is
ambiguous whether the strongest feature is the $8\,\mu{\rm m}$ maximum or
true PAH emission. The physical interpretation of the source, however, 
would be very different for the two alternatives. In order to correctly
identify a feature as a PAH, we require an indication that the 
$6.2\,\mu{\rm m}$ PAH feature is present with the correct shape and (relative to the $7.7\,\mu{\rm m}$ feature) flux. In our sample there is no case of
ambiguity in identifying a feature with the 6.2 or $7.7\,\mu{\rm m}$ PAH feature, since
either both are detected or the large observed width indicates that it cannot be the narrow $6.2\,\mu{\rm m}$ PAH feature. For one of our sources, template fits identify such a single broad maximum with the $7.7\,\mu{\rm m}$ PAH feature but, depending on rest wavelength range and S/N, it is also conceivable to identify such a single broad peak with the $8\,\mu{\rm m}$ maximum of absorbed spectra.

We estimate redshifts by
$\chi ^2$ fitting a selection of templates to the full spectra of the
detected sources. Two of the templates are absorbed, likely AGN dominated, 
sources: NGC 4418 and IRAS F00183$-$7111 \citep{spoon01, spoon04}. As a third
template we use the starburst spectrum of M82 \citep{sturm00}, both as 
observed (with the spectrum dominated by pure PAH emission) and with 
superposition of an additional continuum that is assumed to vary linearly 
over the short wavelength range covered. This last spectrum is similar 
to those of local universe infrared luminous galaxies having significant
contributions to their bolometric luminosities from both star formation 
and powerful AGN (e.g. NGC7469, Mrk273, \citealt{genzel98}). More complex 
fitting schemes have been used to interpret low resolution mid-infrared
spectra, for example by allowing additional obscuration to the PAH 
dominated starburst component \citep[e.g.][]{tran01}, or by decomposing 
the PAH component into individual features that can be well approximated 
by Lorentzians \citep[e.g.][]{boulanger98,smith06}. We do not adopt such
schemes here because of the S/N of our data and their rest wavelength 
coverage, which is usually limited to shortward of the $9.6\,\mu{\rm m}$
silicate absorption by the IRS wavelength range.

Results of the fits with all the templates are listed in Table~\ref{fits}. The best fit for each object is specified in Tab.~\ref{sample} and also shown in Fig.~\ref{fit}. The redshifts derived from the best fit for each object are also listed in Table~\ref{sample} and compared with known redshifts where available. IRS and previous redshifts are consistent for all the sources with previous redshifts from optical spectroscopy (SMMJ09431$+$4700 by \citealt{ledlow02}, SMMJ10519$+$5723, SMMJ10521$+$5719, SMMJ16369$+$4057, SMMJ16371$+$4053 by \citealt{chapman05}) and sometimes from additional CO interferometry (SMMJ02399$-$0136 by \citealt{frayer98}, SMMJ09431$+$4700 by \citealt{neri03,tacconi06}, SMMJ16371$+$4053 by \citealt{greve05}). Three sources had no accurately determined redshifts. For these, our new redshifts are inside the 68\% confidence intervals of the photometric estimates of \citet{aretxaga03} for at least three of the six different evolutionary models used. The formal fit uncertainties for the best fit are up to about $\Delta z=0.02$ but, because of the differences in derived redshifts depending on the template used (Tab.~\ref{fits}), we assume an uncertainty $\Delta z=0.06$ for all the sources. For the six sources with both IRS and optical/CO redshifts measured we derive a reassuringly small standard deviation of only 0.014 for the difference between the best fitting IRS redshift and independent optical/CO redshift. For the uncertainty of a new IRS redshift from a spectrum similar in quality to our spectra, we nevertheless prefer the more conservative $\Delta z=0.06$ which includes the possibility of template mismatch, since the smaller standard deviation for the six sources includes only spectra reasonably well fit by the M82 (plus continuum) templates. $\Delta z=0.06$ is an overestimate if the identification of PAHs in the spectrum is beyond any doubt. Table~\ref{sample} also classifies sources according to how their spectra are characterized: PAH emission (M82 best fit), silicate absorption or $8\,\mu{\rm m}$ maximum (IRAS F00183$-$7111 or NGC4418 best fit) or a superposition of PAH emission and a linearly rising continuum.

In the remainder of this section, we briefly describe the properties of
the nine SMGs with new mid-infrared rest-frame spectra. We also
discuss the four non-detections and constraints on their possible redshifts
and/or SED properties.

{\bf SMMJ00266$+$1708} The best fit for this source is the combination
of M82 PAH template and weak AGN continuum (see Fig.~\ref{fit}), but the 
$\chi ^2$ test gives similar results also for the absorbed continuum
NGC 4418 template (see Tab.~\ref{fits}). The spectrum lacks an evident 
$6.2\,\mu{\rm m}$ PAH feature, the wavelength of which still falls in the
observed band. We proceed with the best fitting PAH plus continuum 
interpretation but note that a classification of this source as an 
absorbed AGN is clearly not excluded. The previous 
photometric redshift of $z=2.7^{+2.3}_{-0.2}$ estimated by \citet{aretxaga03}
is consistent with our spectroscopic value $z=2.73\pm 0.06$. \citet{frayer00} identify a faint ERO counterpart for this object and estimate a lensing magnification $2.4\pm0.5$, which we adopt for our analysis.

{\bf SMMJ02399$-$0136} The mid-infrared spectrum of this source was already
presented by \citet{lutz05}. It contains well detected $6.2\,\mu{\rm m}$ and
$7.7\,\mu{\rm m}$ PAH features superposed on a strong continuum (see Fig.~\ref{fit}).
This source is at the transition between predominantly starburst powered and 
predominantly AGN powered, according to the mid-infrared diagnostics of
\citet{genzel98} and \citet{laurent00}. AGN signatures are also seen in 
the optical spectrum \citep{ivison98}. Combining mid-infrared spectroscopy
and \textit{Chandra} X-ray observations \citep{bautz00}, we conclude that
this source is powered by roughly equal contributions of star formation and 
a Compton-thick AGN \citep{lutz05}, in agreement with constraints on the
importance of star formation in this object from molecular gas mass
and its position relative to the radio/far-infrared correlation
\citep{frayer98}. Its luminosity (see Tab.~\ref{sample}) 
is calculated taking into account a lensing magnification factor of
2.5 \citep{ivison98}.

{\bf SMMJ09429$+$4659} This source is well fitted with the starburst-like M82 spectrum. No accurate redshift was previously known for this object. Using the radio-submm spectral index and the models from \citet{carilli00}, the estimated redshift is $z=0.4\pm 0.3$. The intense radio emission suggests, though, that there is a significant radio contribution from an AGN, even if we do not see AGN emission in the mid-infrared spectrum, making the value estimated from the radio-submm relation only a lower limit. Assuming this galaxy follows the $K-z$ relationship for powerful radio galaxies \citep{jarvis01}, it most likely lies at $z\geq 2$ \citep{ledlow02}. Our spectroscopic value of $z=2.38\pm 0.06$ confirms the high redshift of this object. The luminosity of this source (see Tab.~\ref{sample}) is calculated by taking into account a lensing magnification factor of 1.3 \citep{cowie02}.

{\bf SMMJ09431$+$4700} We pointed IRS at the mm position \citep{neri03}
of component H7 in the notation of \citet{ledlow02}, but note that
component H6 is included in the observing aperture as well. This source 
has the highest redshift of the sample ($z=3.36$), not accounting for the
non-detected objects (see discussion below). It is well fitted with a PAH
spectrum plus a strong linearly rising continuum, so we infer that it 
is powered by both starburst activity and a powerful AGN (see Fig.~\ref{fit}).
The optical spectrum of component H6 shows features of a weak AGN 
\citep{ledlow02}. The spectral properties, line widths and line ratios 
of this galaxy are very similar to those seen for narrow-line Seyfert 1
galaxies (NLSy1s; \citealt{crenshaw91}). \textit{XMM-Newton}
observations \citep{ledlow02} suggest that the intrinsic
X-ray luminosity of the AGN is modest
(L$_{\rm 2-10\,keV}\lesssim 10^{44}\,{\rm erg\,\,s}^{-1}\,\mathrm{cm}^{-2})$, unless there
is heavy obscuration. Comparison to the strong rest frame mid-IR
continuum seen in the IRS spectrum suggests the latter is indeed the case: at
L$_{\rm 2-10\,keV}/\nu \rm{L}_\nu (6\,\mu \rm{m})\lesssim 0.015$, this source falls more than an order of magnitude below the relation between unobscured $2-10\,{\rm keV}$
luminosity and $6\,\mu{\rm m}$ continuum for local AGN \citep{lutz04}. SMMJ09431$+$4700
hosts a heavily obscured or Compton-thick AGN, the location
of which we cannot firmly ascribe to component H6 or H7. The luminosity of 
SMMJ09431$+$4700 (see Tab.~\ref{sample}) is calculated
taking into account a lensing amplification factor of 1.2 \citep{cowie02}.

{\bf SMMJ10519$+$5723} The lowest $\chi ^2$ fit for this object is the one
with the M82 starburst template, which is supported by the tentative detection of
a $6.2\,\mu{\rm m}$ PAH feature. Still, the $\chi ^2$ is not much higher for
the absorbed continuum template NGC 4418. We interpret this
source as powered by starburst activity, consistent with optical 
spectroscopy \citep{chapman05}, but note that an absorbed continuum
interpretation cannot be firmly excluded. The spectroscopic redshift 
suggested for this source by \citet{chapman03} was $z=3.699$. Later,
\citet{egami04} and \citet{chapman05} indicated values of $z=2.69$
and $z=2.686$ respectively. Our fits give  $z=2.67\pm 0.06$ using
the starburst template and $z=2.64\pm 0.06$ using the obscured one.
Both redshifts are in agreement with the latest optical results.

{\bf SMMJ10521$+$5719} The spectrum of this source shows PAH features
plus a weak continuum. The source does not contain AGN features in the optical spectrum \citep{chapman05}. The mid-infrared AGN continuum is
detected but the feature-to-continuum ratio is much higher than in the
cases of SMMJ02399$-$0136 and SMMJ09431$+$4700, thus suggesting a smaller
AGN contribution. The presence of an AGN is confirmed as well from radio 
and X-ray emission that show the possible presence of a radio loud
quasar \citep{ivison02}.

{\bf MMJ154127$+$6616} The mid-infrared spectrum of this source was
already published by \citet{lutz05}. We reprocessed the data using a 
later version of the IRS pipeline (14.0.0). The new reduction further
increases the similarity to a starburst spectrum with well defined 
$6.2\,\mu{\rm m}$ and $7.7\,\mu{\rm m}$ PAH features and is well fitted by a M82
spectrum plus a very weak flat continuum. The $\chi ^2$ is not much worse
adopting the NGC4418 template which better 
matches the emission at long wavelengths, but the clear presence of both 
PAH features strongly increases confidence that this source is powered by 
star formation.

{\bf SMMJ16369$+$4057} This source has the lowest redshift of the sample
($z=1.21$) and is well fitted with the M82 spectrum plus a very weak flat
continuum. The optical spectroscopy of \citet{chapman05} detects typical 
starburst lines, consistent with our result.

{\bf SMMJ16371$+$4053} The mid-infrared spectrum of this source shows
clear $6.2\,\mu{\rm m}$ and $7.7\,\mu{\rm m}$ PAH features. A flat continuum is
detected. The mid-infrared spectrum shows that it is powered mainly by
starburst activity, even though an AGN is probably also present. AGN lines
have been detected in  the optical spectrum of \citet{chapman05}.

{\bf IRS non-detections} The sources for which we have neither detected
features nor continua in the IRS spectra despite accurate
interferometric positions are
MMJ120517$-$0743.1, MMJ120539$-$0745.4, MMJ120546$-$0741.5 and MMJ154127$+$6615.
Multiple arguments support the reality of these mm sources: they are well
detected in the original MAMBO data 
\citep{bertoldi00,dannerbauer02,dannerbauer04a}, are confirmed by SCUBA
\citep{eales03}, have weak VLA counterparts
\citep{bertoldi00,dannerbauer04a}
and have dust continuum emission directly confirmed and located by mm
interferometry \citep{dannerbauer02,dannerbauer04a,dannerbauer04b}. The
faintness or non-detection of their optical/near infrared counterparts
(see also Table 3) lead these authors to the conclusion that these sources must
be at very high redshift and/or highly obscured.

We now consider each of these scenarios in turn. If the IRS non-detections are intrinsically similar to the detected sources but simply more distant, we can place one lower limit on their redshifts by requiring that the $7.7-8.6\,\mu{\rm m}$ PAH complex or the $8\,{\rm \mu m}$ maxima in any absorbed spectra have remained undetected because they were shifted to wavelengths longer than $\sim 35\,{\rm \mu m}$, where the detector sensitivity
falls rapidly and the noise increases. This argument implies redshifts greater than 3.6. Similarly, since $6.2\,{\rm \mu m}$ features in PAH emission spectra would already be lost in the noise at wavelengths longer than $\sim 30\,{\rm \mu m}$ (being somewhat weaker), such sources could only be IRS non-detections for $z > 3.8$. Hence, we would conclude for all undetected sources redshifts $\gtrsim 3.6$.

If the IRS non-detections lie at redshifts similar to those of the IRS detections, their SEDs must be significantly different. We infer in Section 5.2 below that the detected sources already have a ratio of mid-infrared PAH to far-infrared emission similar to that of local ULIRGs.  This implies that to escape IRS detection at the same redshifts, the non-detections would need to be even more heavily obscured in the mid-infrared than local ULIRGs. By way of example, consider Arp 220, well known for its extremely low ratio of mid- to far-infrared emission \citep[e.g.][]{sanders88,haas01}. At $z \sim 2.5$, an extreme Arp 220-like SED scaled to the millimeter fluxes of the IRS non-detections would still manifest a broad PAH feature peaking at the $0.15-0.2\,{\rm mJy}$ level. No such feature is indicated in the spectra of the four non-detections (Fig.~\ref{fit}).

Previous assessments of the ``higher redshift'' and ``higher obscuration'' scenarios for these sources have been based on faint or undetected near-IR counterparts. For MMJ120517--0743.1, MMJ120539--0745.4, and MMJ120546--0741.5, faint $K_s$-band magnitudes and the assumption of SEDs similar to those of local ULIRGs imply very high redshifts $z \gtrsim 4$ \citep{dannerbauer02}. Similarly, the $K > 21.2$ counterpart of MMJ154127$+$6615 implies a redshift $z \gtrsim 3$.  This particular argument is weakened by evidence that SMGs can have rest-frame {\it UV/optical} obscurations greater than those of local ULIRGs; SMMJ00266$+$1708, for example, has a very faint near-IR counterpart but (based on our data) $z = 2.73$. However, if we generalize the argument to the rest-frame {\it mid-infrared}, which is more difficult to obscure than the UV/optical, we are on stronger ground. Tellingly, our IRS non-detections are at least $\sim 3$ times fainter in the rest-frame mid-infrared than even SMMJ00266$+$1708.  This result is confirmed by the mid-infrared imaging of \citet{charmandaris04}, who place all four sources (three not detected and one tentatively detected) at the low end of the mid-IR to submm flux ratio distribution.

A further reason to prefer the ``higher redshift'' scenario for the IRS non-detections is that the radio counterparts to all four are faint, despite their bright mm fluxes \citep{bertoldi00,dannerbauer04a}. The mean of their ratio of $850\,{\rm \mu m}$ and $1.4\,{\rm GHz}$ flux densities ($212\pm 77$) is about twice that of the \citet{chapman05} sample ($95\pm 9$), and still higher than both the value of $108\pm 9$, obtained for the same sample after exclusion of potentially radio-loud ($S_{\rm 1.4\,GHz} > 200\,{\rm \mu Jy}$) SMGs, and the value of $150\pm 20$, derived from a 15 sources subsample matching the mean SCUBA flux of our four non-detections. This trend is consistent with higher redshifts \citep{carilli00}, although the scatter about the radio/submm vs. $z$ relation is large. Taken together, the various lines of evidence suggest that the four IRS non-detections do lie at high redshift, although with all arguments still based on SED assumptions. Direct spectroscopic redshifts will be needed for a definitive conclusion.

\section{Discussion}

\subsection{The redshift distribution has a median of $z\sim 2.8$}

\begin{figure}[p!]
\centering

\includegraphics[width=13.cm,keepaspectratio]{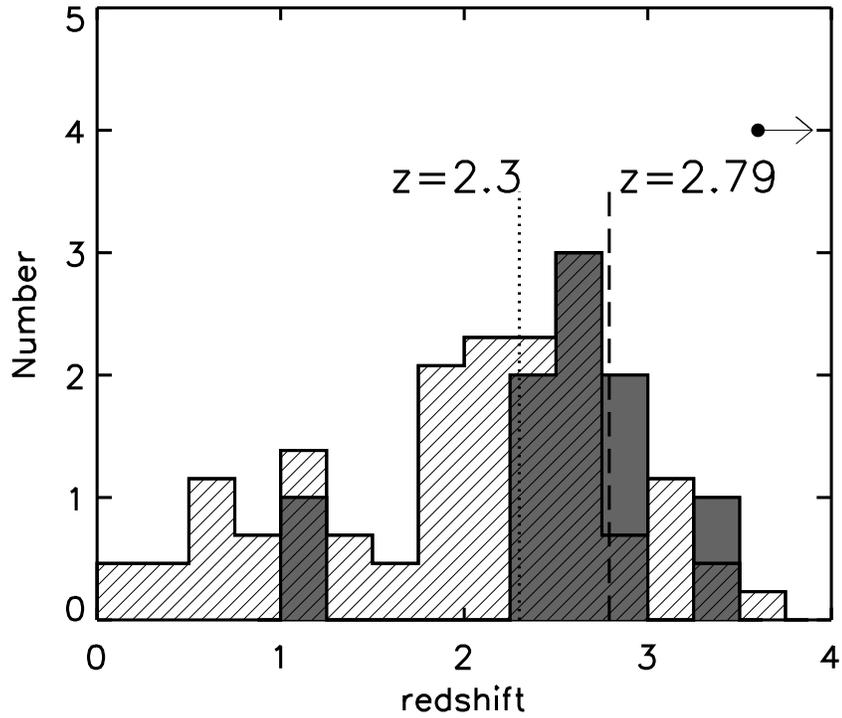}
\caption{Histogram showing the redshift distribution for our SMG sample (solid) and for the submillimeter-flux-limited sample of \citet{chapman05} (cross-hatched), scaled for the maximum value. The dotted line at $z=2.3$ indicates the median value obtained by \citet{chapman05} and the dashed line at $z=2.79$ indicates the median of our sample distribution. Our value is calculated assuming $z\geq 3.6$ for undetected sources.}
\label{redshifts}
\end{figure}

Figure~\ref{redshifts} shows the redshift distribution for our sample.
Taking into account the lower limits adopted for the undetected sources,
we derive a median redshift of $z=2.79$ for the full sample of 13 SMGs.
The median redshift of the 9 detected sources is $z=2.69$. These values
are noticeably higher than the median $z=2.2$ measured by \citet{chapman05}
for their sample of 73 submillimeter galaxies with optical redshifts, and
also higher than their estimate of $z=2.3$ for the extrapolation to the
full SMG population. None of our new IRS redshifts is in the optical
``redshift gap'' ($1.2\lesssim z\lesssim 1.8$) of the optical spectroscopic
census of the submillimeter galaxy population \citep{chapman05}, where
some new IRS redshifts might have been expected. It is natural to assume that 
the optical redshifts are still biased towards the optically bright and
low redshift part of the population \citep{dannerbauer04a}. The recent 
discovery of a submillimeter galaxy at redshift $z\sim 4$ \citep{knudsen06}
similarly indicates an extension of the SMG redshift distribution beyond 
the one established by \citet{chapman05}. In contrast, \citet{pope06} 
suggest from a combination of spectroscopic and photometric redshifts 
a median redshift of 2.0 for those SMGs in the HDF-North region that
they consider securely identified through radio or \textit{Spitzer} counterparts.

Our results suggest a modest extension of the \citet{chapman05} redshift 
distribution towards a larger high redshift tail, but the number
statistics of our and other current SMG samples with redshifts is small.
To investigate the significance of this difference, we have run simple 
Monte-Carlo simulations to estimate the probability of obtaining by chance a median redshift $\geq 2.79$ for a 13 objects sample drawn randomly from a \citet{chapman05} redshift distribution.
Using the distributions plotted in Fig.~4 of \citet{chapman05},
this probability is a low 0.3\% drawing 13-objects samples from their
optical SMG redshift distribution with overall median 2.2, and a still
interesting 8\% for drawing from their suggested
extrapolation to the overall SMG redshift distribution with median 2.3.
The latter comparison is conservative in ignoring the fact that our
sample with mostly {\it radio}-interferometric identifications is still
biased against the very highest redshift objects.  

In addition to these simple statistical comparisons, and perhaps more
important, there can be effects of field-to-field variations including
spikes in the redshift distribution for the current small area SMG surveys
\citep{blain04}. A potential $z>4$ redshift spike in the NDF region
observed with MAMBO by \citet{dannerbauer04a} from which three of our
IRS non-detections are drawn, for example, could emphasize
the high z component in our small sample. Likewise, the results of
\citet{pope06} could have been driven in the opposite direction by being drawn from a single field. These constraints clearly call for further analyses using larger samples from large and widely separated fields.

\subsection{SMGs have ULIRG-like SEDs and are largely starburst-powered}

\begin{figure}[p!]
\centering

\includegraphics[width=13.cm,keepaspectratio]{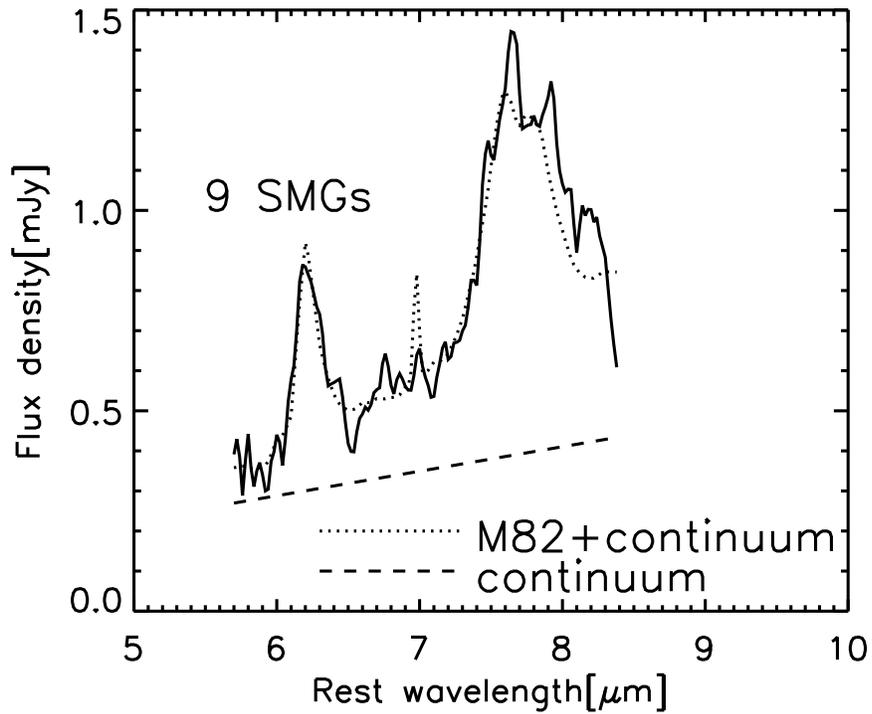}
\caption{Average IRS spectrum of all 9 detected SMGs, individually scaled to  $S_{222\,\mu \mathrm{m}}=15\,{\rm mJy}$ in the rest wavelength. Both the $6.2\,\mu{\rm m}$ and the $7.7\,\mu{\rm m}$ PAH features are clearly visible. The spectrum is well fitted by the starburst-like spectrum of M82 plus a weak continuum.}
\label{average}
\end{figure}

The rest frame mid-infrared spectra of SMGs and their comparison to the
far-infrared part of the SEDs hold interesting clues
about their physical properties, energy sources and radiation fields.
The average spectrum of the nine detected SMGs, individually scaled to the
same rest wavelength flux $S_{222\,\mu \mathrm{m}} = 15\,{\rm mJy}$ to give all
sources equal weight (see Fig.~\ref{average}), clearly shows the PAH
features at 6.2 and $7.7\,\mu{\rm m}$ but relatively weak continuum. The
rest wavelength of $222\,\mu{\rm m}$ was chosen for this scaling because it is
well constrained by observations, since it is the rest frame wavelength
for the SCUBA $850\,\mu{\rm m}$ flux at redshift $z=2.8$, about the median redshift
of the sample. Scaling by the rest frame far-infrared emission is close
to scaling by bolometric flux, which is appropriate for interpreting
the mid-infrared diagnostics in Fig.~\ref{average} in terms of energy sources
of a typical SMG without biasing towards star formation (PAH) or AGN
(continuum). For galaxies at different redshifts, the rest frame $222\,\mu{\rm m}$
continuum emission was extrapolated from the SCUBA flux.
For this step as well as in quantifying the bulk properties of our targets,
we assign luminosities and temperatures following the approach of
\citet{chapman05} who use the local FIR/radio relation \citep{helou85}
to assign FIR SEDs to sources with known redshift, radio, and submm fluxes.
The adopted far-infrared SED model is a grey body with
$S_{\nu,T}\propto k(\nu)B_{\nu,T}$ with $k(\nu)\propto \nu^{\beta}$
and $\beta =1.5$.
We calculated $T_\mathrm{d}$ for our galaxies using two photometric points ($850\,\mu{\rm m}$ and $1.4\,{\rm GHz}$) and the relation found by \citet{chapman05}
\begin{equation}
T_\mathrm{d}\propto\frac{1+z}{(S_{850\,\mu \mathrm{m}}/S_{1.4\,\mathrm{GHz}})^{0.26}}
\end{equation}
with a proportionality constant of 6.29 (see Tab.~\ref{sample}). Infrared luminosities are calculated as the integral between 8 and $1000\,\mu{\rm m}$ of the SED, assuming a $\Lambda$CDM cosmology with $H_0=70\,{\rm km s}^{-1}{\rm Mpc}^{-1}$,
$\Omega_M = 0.3$ and $\Omega_{\Lambda} = 0.7$ (see Tab.~\ref{sample}).
Like other SMGs, our sample objects are inferred to be very luminous objects
($L_{\rm IR} \sim 10^{13}L_{\bigodot}$).

\begin{figure}[p!]
\centering

\includegraphics[width=13.cm,keepaspectratio]{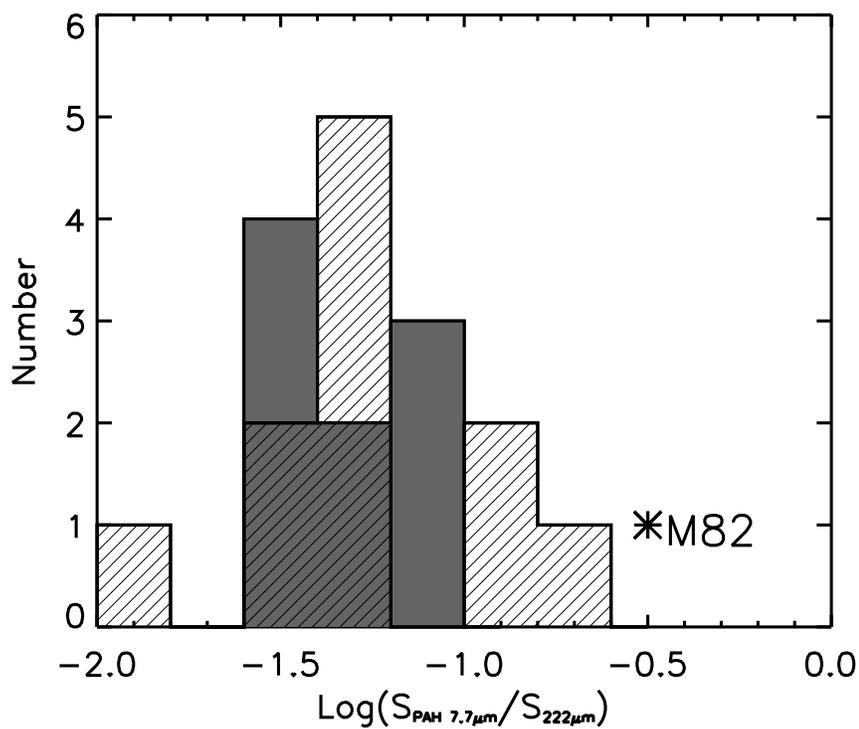}
\caption{Histogram showing the ratio of PAH $7.7\,\mu{\rm m}$ peak flux density and rest frame $222\,\mu{\rm m}$ continuum flux density for our SMG sample (solid) and for eleven local ULIRGs (cross-hatched). In this measurement of the mid-to far-infrared SED, SMGs are very similar to the local ULIRG population. However, they show a lower value than the low luminosity starburst M82. The M82 point is based on the PAH data of \citet{foerster03} and the far-infrared continuum of \citet{colbert99}, obtained in large and similar apertures.}
\label{histogram}
\end{figure}

The spectrum of Fig.~\ref{average} provides a first and direct
indication that our sources are,
on average, starburst-like \citep[see also][for a similar conclusion for
several lower redshift SMGs]{menendez07}. It can be seen
as a superposition of a M82-type
PAH spectrum and a weak additional continuum. Comparison with the spectra
of the mostly starburst-dominated local ULIRG population (see e.g. Fig.~1
in \citealt{lutz98}) suggests that SMGs are scaled up versions of these
objects. A further proof of this comes from a comparison of the ratio of
PAH features and $S_{222\,\mu \mathrm{m}}$ continua. Figure~\ref{histogram}
shows the ratio of peak flux density of the $7.7\,\mu{\rm m}$ PAH feature, after
continuum subtraction, to the continuum flux density at $222\,\mu{\rm m}$ rest
frame for all the detected sources of our sample and for 11 local ULIRGs 
with PAH emission and good FIR photometry. The ULIRG data have been 
taken from ISOPHOT-S
observations \citep{rigopoulou99} and the continua from slight 
extrapolations of the far-infrared photometry of \citet{klaas01}, which
extends to an observed wavelength of $200\,\mu{\rm m}$. The PAH-to-far-infrared
ratios of our sources are fully consistent with those of the local ULIRG 
population. These SED properties are in agreement with the conclusion, 
from spatially resolved mm interferometry, that SMGs are similar to
local ULIRGs suitably scaled for their larger masses, luminosities and 
star formation rates, as well as their greater gas fractions \citep{tacconi06}.

Given this similarity of SMGs and local ULIRGs in the comparison between 
PAH and far-infrared parts of the SED, it is instructive to also compare the
dust temperatures for $z\approx 2.5$ SMGs with those of local galaxies.
\citet{chapman05} infer $T_\mathrm{d}\sim 36\pm 7$K for SMGs which can be compared
to local galaxies in two ways: (1) SMGs have colder dust temperatures than 
local ULIRGs of the same luminosity (that means colder dust than so-called
HyLIRGs). For local HyLIRGs, \citet{chapman03b} find $T_\mathrm{d}\sim 42$K at $L \sim 10^{13}L_{\bigodot}$. This could be due to their difference in energy sources: local HyLIRGs seem predominantly AGN driven \citep[e.g.][]{tran01}, while many SMGs of the same luminosity are starbursts. (2) SMGs have similar dust temperatures to the bulk of the local ULIRG population which has luminosities just above $10^{12}L_{\bigodot}$. For local ULIRGs, \citet{chapman03} infer
$T_\mathrm{d}\sim 36$K. This comparison is in line with other indications that SMGs are ``scaled up'' versions of such ULIRGs rather than direct analogs of local HyLIRGs.

From the rest frame mid-infrared spectra it is possible to evaluate the
presence and the strength of a possible AGN contribution to the very 
large IR luminosity of the SMGs of our sample. Many SMGs show evidence 
of an AGN, from X-ray observations
\citep{alexander03,alexander05,alexander05b} or optical/near-infrared
spectra \citep[e.g.][]{ivison98,swinbank04,chapman05}. However, the most 
important question is not whether 
there are detectable AGN signatures, but whether or not the AGN is a 
significant contributor to the luminosity of the galaxies. From our spectral 
decompositions, we can use the ratio of PAH $7.7\,\mu{\rm m}$ peak and local continuum
as an indicator of the AGN and star formation contribution to the bolometric
luminosity. Following studies of local infrared galaxies 
\citep{genzel98,laurent00,tran01}, we adopt a feature to continuum ratio of 1
as the border between predominantly star formation and predominantly AGN 
powered\footnote{Our decomposition by template fit differs from the 
one used by \citet{genzel98}, which interpolates between two points in the
observed spectrum to define the continuum, in approaching very high 
feature to continuum values in the limit of almost pure star formation. The
two methods give very similar results in the presence of significant AGN
continuum}. The fitted fluxes for the $7.7\,\mu{\rm m}$ PAH peak and local
continuum are listed in Table~\ref{sample}.

In our sample, there is no trace of an AGN continuum in the adopted fits
to the rest frame mid-infrared spectrum for two sources (SMMJ09429$+$4659
and SMMJ10519$+$5723).
Seven sources (SMMJ00266$+$1708, SMMJ02399$-$0136, SMMJ09431$+$4700,
SMMJ10521$+$5719,\\ MMJ154127$+$6616, SMMJ16369$+$4057, SMMJ16371$+$4053)
are well fitted by a superposition of an AGN continuum and PAH features.
We note again that the alternative fit by an obscured (AGN?) continuum cannot
be firmly excluded for one of these (SMMJ00266$+$1708). Assuming that a feature to continuum ratio of 1 means similar bolometric contribution from
star formation and accretion, we conclude that, of these nine targets, two are pure starbursts with at best very weak AGN, five have AGN with modest contributions
of the order 20\%, and two (SMMJ02399$-$0136, SMMJ09431$+$4700) have strong AGN
contributing slightly above half of the bolometric luminosity.   
We cannot constrain the starburst or AGN nature of the four undetected 
sources that are likely at high redshift, apart from stating that their
non-detection means absence of a strong unobscured hot AGN continuum
even for redshifts somewhat above 3.6. Our sample is small, so
it seems reasonable to expect that the full submillimeter population will
show an even wider range of AGN properties perhaps including less obscured, energetically dominant AGNs. \citet{egami04} and \citet{ivison04} have used mid-IR
photometry as a diagnostic tool to put limits on the AGN contribution
to the IR luminosity and agree that fewer than $25\%$ of the SCUBA/MAMBO
sources observed are AGN powered. From X-ray observations,
\citet{alexander05b} found that, on average, the AGN contribution in a 
large sample of radio-detected SMGs was likely to be modest
($\simeq 10\%$). This result assumes that SMGs do not have a substantially
larger dust-covering factor than optically selected quasars, and that there
is no significant number of fully Compton-thick AGNs that are hard to
detect in X-rays. Our finding of starbursts being prevalent in the
mid-infrared spectra extends this result by showing that SMGs
typically do not contain such dominant X-ray obscured AGNs with strong
mid-infrared continuum re-emission. Obscured AGNs with strong
mid-infrared continuum are found in \textit{Spitzer} mid-infrared
surveys \citep[e.g.][]{martinez05}, but show only little overlap with the 
SMG population \citep{lutz05b}. In general it seems that the AGN contribution 
to the IR luminosity of most SMGs is small when compared to heating from 
star-formation activity.

While the IRS spectra argue against dominant AGN being
typical for SMGs, they can still help to constrain the obscuration of 
the lesser AGN found, independent of whether sufficient photons are
available for a detailed X-ray spectral analysis \citep{alexander05b}.
From comparison to X-ray data, we have argued in the discussion of the
individual sources for high X-ray obscuration of the two strongest AGN in our
sample. For $z=2.8$, a rest frame $6\,\mu{\rm m}$ continuum of $\sim0.1\,{\rm mJy}$ for the weaker AGN, and the relation
of rest frame $2-10\,{\rm keV}$ flux and $6\,\mu{\rm m}$ AGN continuum from \citet{lutz04},
a rest frame $2-10\,{\rm keV}$ emission of $4\times 10^{-15}{\rm erg\,\,s}^{-1}{\rm cm}^{-2}$ is
expected for an unobscured AGN following this relation. This is in the
sensitivity regime of current X-ray data in the corresponding observed
frame band, and can constrain the obscuration of such ``minor'' AGN with the
caveat that large samples are needed given the variations in intrinsic
AGN SEDs and the corresponding significant scatter of the
mid-IR/X-ray relation.
We have used \textit{Chandra} archival X-ray data to put limits on the observed frame $0.5-2\,{\rm keV}$ emission of SMMJ00266$+$1708 ($<0.56\times 10^{-15}
{\rm erg\,\,s}^{-1}{\rm cm}^{-2}$) and MMJ154127$+$6616 ($<0.16\times 10^{-15}$). From \textit{XMM-Newton} observations of SMMJ10521$+$5719 in the same soft band, a value of $0.16\times 10^{-15}{\rm erg\,\,s}^{-1}{\rm cm}^{-2}$ is derived (Brunner et al., in prep.). These limits and fluxes imply that the
rest frame $2-10\,{\rm keV}$ emission is lower than the extrapolation from the
mid-infrared continuum to unobscured X-rays. Noticeable X-ray
obscuration may thus be found in many of the minor AGN in SMGs, in 
agreement with \citet{alexander05b}.

Metallicity and dust-to-gas ratios in many high redshift galaxies are
expected to be lower than at low redshift. Low metallicity systems show 
weaker mid-infrared PAH emission bands \citep[e.g.][]{engelbracht05}.
In addition to differences in radiation fields, this is probably due to 
the fact that these galaxies are young and thus may lack the carbon-rich
AGB stars required to form the PAH molecules. However, the enrichment 
will proceed once intense star formation activity is underway for a 
sufficient time, or if there has been preceding star formation. This
seems to be the case for SMGs. They show very clear PAH features as 
tracers of their intense starbursts. We can therefore assume that high 
metallicity is typical for the population of massive SMGs, consistent 
with the supersolar metallicity derived from nebular emission for 
SMM14011 \citep{tecza04} and the roughly solar abundances in the SMG 
sample of \citet{swinbank04}. In fact, because of the intense star 
formation, the metallicity of these systems should rapidly approach
that of their likely present-day descendants: luminous
elliptical galaxies \citep{swinbank04}. These observations and the 
scenario that SMGs evolve to ellipticals are in full agreement with 
the fossil record that the formation of the stars of local ellipticals 
must have happened rapidly and at high redshift \citep{thomas05}.

\section{Conclusions}

We have presented \textit{Spitzer} mid-infrared spectra of a sample of
13 submillimeter galaxies. For nine of them, we have unambiguous 
detections of PAH spectral features and/or mid-infrared continua
that allow us to constrain energy sources in these objects and to 
determine, in three cases for the first time, their redshifts.

The IRS detections alone have a median $z \sim 2.7$.  If the four IRS non-detections lie at similar redshifts, their rest frame mid-infrared obscurations would have to be even more extreme than those of local ULIRGs. More plausibly, the four IRS non-detections lie at higher redshifts ($\geq 3.6$), giving a median $z \sim 2.8$ for the full set of 13.  Although our sample is small, this result may indicate an extension to higher redshift of the SMG redshift distribution relative to radio-preselected samples with optical redshifts.

In the majority of cases, the detection of PAH emission and the
weakness of AGN continua indicate that these galaxies are mainly 
starburst-powered. This result agrees with previous X-ray, optical
and SED studies that indicate only a small AGN contribution to the 
IR luminosity compared to heating from star-formation activity. Our
work extends these studies by also constraining the role of highly 
obscured AGN.

The SED properties of our galaxies are in agreement with the SMGs 
being scaled-up versions of the compact star-forming regions in local ULIRGs.

The existence of star formation dominated systems at infrared luminosities in excess of $10^{13}L_{\bigodot}$ is unique to the high redshift universe. The presence of high luminosity starbursts in SMGs may be related to their higher gas fractions \citep{greve05,tacconi06}.

Mid-infrared spectroscopy with IRS, together with ancillary observations from the optical through radio wavelengths, can play a central role in understanding the nature of submillimeter galaxies and can be a powerful tool for probing the earliest and most dramatic stage of the evolution of galaxies.

\vspace{1cm}
This work is based on observations made with the \textit{Spitzer Space Telescope}, which is operated by the Jet Propulsion Laboratory, California Institute of Technology, under a contract with NASA. We thank the referee for thorough and helpful comments.
The authors want to thank M.Brusa and V.Mainieri for help with X-ray data.
E.V. would like to thank the following people for helpful discussions and support during the work: M.Righi, F.Braglia, L.Conversi.


\clearpage

\begin{deluxetable}{l l r r c c c c}
\rotate
\tablecolumns{8}
\tablewidth{42pc}
\tablenum{1}
\tabletypesize{\scriptsize}
\tablecaption{Summary of SMG observations}
\tablehead{
\colhead{Name} & \colhead{RA} & \colhead{DEC} & \colhead{Error} & \colhead{Ref.} &\colhead{SL1} & \colhead{LL2} & \colhead{LL1} \\
\colhead{} & \colhead{J2000} & \colhead{J2000} & \colhead{\arcsec} & \colhead{} & \colhead{($7.4-14.5\,\mu{\rm m}$)} & \colhead{($14.0-21.3\,\mu{\rm m}$)} & \colhead{($19.5-38.0\,\mu{\rm m}$)} \\
\colhead{} &\colhead{} &\colhead{} &\colhead{} &\colhead{} & \colhead{s$\times$n.of cycles} & \colhead{s$\times$n.of cycles} & \colhead{s$\times$n.of cycles}
}
\startdata
SMMJ00266$+$1708 (M12)     &00:26:34.10&$+$17:08:33.7&0.8&F00& & $120\times 15$  & $120\times 30$     \\
SMMJ02399--0136 (L1/L2)  &02:39:51.87&$-$01:35:58.8&0.6&G03& &                 & $120\times 30$ \\
SMMJ09429$+$4659  (H8)     &09:42:53.42&$+$46:59:54.5&&L02& & $120\times 15$  & $120\times 30$     \\
SMMJ09431$+$4700 (H7/H6)   &09:43:03.69&$+$47:00:15.5&0.3&N03& &                 & $120\times 30$  \\
SMMJ10519$+$5723 (LE\,850.18)&10:51:55.47&$+$57:23:12.7&&I02& &                 & $120\times 30$  \\
SMMJ10521$+$5719 (LE\,850.12)&10:52:07.49&$+$57:19:04.0&&I02& &                 & $120\times 30$ \\
MMJ120517--0743.1        &12:05:17.86&$-$07:43:08.5&0.4&D04& & $120\times 10$  & $120\times 30$      \\
MMJ120539--0745.4        &12:05:39.47&$-$07:45:27.0&0.4&D04& &                 & $120\times 30$      \\
MMJ120546--0741.5        &12:05:46.59&$-$07:41:34.3&0.5&D04& & $120\times 11$  & $120\times 30$     \\
MMJ154127$+$6615           &15:41:26.90&$+$66:14:37.3&0.1&B00& & $120\times 15$  & $120\times 30$      \\
MMJ154127$+$6616           &15:41:27.28&$+$66:16:17.0&0.1&B00& & $120\times 15$  & $120\times 30$      \\
SMMJ16369$+$4057 (N2\,850.8) &16:36:58.78&$+$40:57:28.1&&I02& $240\times 5$  & $120\times 10$   &  \\
SMMJ16371$+$4053 (N2\,1200.17)&16:37:06.60&$+$40:53:14.0&&G05& & $120\times 15$  & $120\times 30$  \\
\enddata
\tablecomments{Names in brackets indicate aliases used in the literature.}
\tablerefs{Interferometric positions adopted for the IRS observations (from 
VLA $1.4\,{\rm GHz}$ if not stated otherwise):
B00: \citet{bertoldi00}. 
D04: PdBI mm position of \citet{dannerbauer04a}.
F00: OVRO mm position of \citet{frayer00}.
G03: PdBI mm position of \citet{genzel03}.
G05: 1\arcsec\ from final CO position of \citet{greve05}.
I02: \citet{ivison02}.
L02: \citet{ledlow02}.
N03: PdBI mm position of \citet{neri03}. For sources with no position error stated in the references, we assume $\lesssim 0.5\arcsec$ from the interferometer configurations used.
}
\label{observations}
\end{deluxetable}

\clearpage

\begin{deluxetable}{l c c c c c c c c c c c c c c c c}
\tablecolumns{17}
\tablewidth{39pc}
\tabletypesize{\scriptsize}
\tablenum{2}
\tablecaption{Fit results. The best fit is in bold type (see the text for details).}
\tablehead{
\colhead{Name} & \colhead{} &
\multicolumn{3}{c}{M82} & \colhead{} &
\multicolumn{3}{c}{IRAS F00183$-$7111} & \colhead{} &
\multicolumn{3}{c}{NGC 4418} & \colhead{} &
\multicolumn{3}{c}{M82 $+$ continuum} \\
\cline{3-5}
\cline{7-9}
\cline{11-13}
\cline{15-17}
\colhead{}     & \colhead{} &
\colhead{$\chi^{2}$} & \colhead{$z$} & \colhead{$\sigma_{z}$} & \colhead{} &
\colhead{$\chi^{2}$} & \colhead{$z$} & \colhead{$\sigma_{z}$} & \colhead{} &
\colhead{$\chi^{2}$} & \colhead{$z$} & \colhead{$\sigma_{z}$} & \colhead{} &
\colhead{$\chi^{2}$} & \colhead{$z$} & \colhead{$\sigma_{z}$}
}
\startdata
SMMJ00266$+$1708     & &1.5 &2.73&0.01& &3.1 &2.75&0.02& &2.0 &2.66&0.01& &
\textbf{1.4} &2.73&0.02\\
SMMJ02399$-$0136     & &17  &2.80&0.01& &12  &2.86&0.01& &21  &2.72&0.01& &
\textbf{1.1} &2.81&0.02\\
SMMJ09429$+$4659     & &\textbf{0.58}&2.38&0.01& & 1.5 &2.43&0.03& &0.86&2.33&0.01& &
0.59&2.38&0.02\\
SMMJ09431$+$4700     & &10  &3.33&0.01& &3.2 &3.26&0.04& &4.3 &3.20&0.01& &
\textbf{0.99}&3.36&0.02\\
SMMJ10519$+$5723     & &\textbf{0.68}&2.67&0.01& &1.2&2.67&0.01& &0.80&2.64&0.02& &
0.75&2.67&0.02\\
SMMJ10521$+$5719     & &0.74&2.69&0.01& &1.3 &2.75&0.02& &1.1 &2.73&0.01& &
\textbf{0.63}&2.69&0.02\\
MMJ154127$+$6616     & &0.55&2.78&0.01& &0.82&2.79&0.04& &0.60&2.67&0.02& &
\textbf{0.35}&2.79&0.02\\
SMMJ16369$+$4057     & &0.60&1.21&0.01& &1.68&1.17&0.01& &0.66&1.12&0.01& &
\textbf{0.30}&1.21&0.02\\
SMMJ16371$+$4053& &1.1 &2.38&0.01& &1.2 &2.48&0.01& &1.3 &2.34&0.01& &
\textbf{0.41}&2.38&0.02\\
\enddata
\label{fits}
\end{deluxetable}

\clearpage

\begin{deluxetable}{l r r r r c c c c c c c c}
\rotate
\tablecolumns{13}
\tablewidth{50pc}
\tablenum{3}
\tabletypesize{\scriptsize}
\tablecaption{Properties of the SMG sample}
\tablehead{
\colhead{Name} &
\colhead{S$_{850\,\mu \mathrm{m}}$} &
\colhead{S$_{1.4\,\mathrm{GHz}}$} &
\colhead{$K$} &
\colhead{Opt.} & 
\colhead{Magn.\tablenotemark{a}} &
\colhead{$z$\tablenotemark{b}} &
\colhead{$z_{\mathrm{{f}}}$\tablenotemark{c}} &
\colhead{S$_{\mathrm{PAH} 7.7\,\mu \mathrm{m}}$\tablenotemark{d}} &
\colhead{S$_{\mathrm{cont.} 7.7\,\mu \mathrm{m}}$\tablenotemark{e}} &
\colhead{$T_{\mathrm{d}}$} &
\colhead{$L_{\mathrm{IR}}\times 10^{13}$} &
\colhead{Best fit} \\
\colhead{} &
\colhead{mJy} &
\colhead{$\mu{\rm Jy}$} &
\colhead{mag} &
\colhead{mag} &
\colhead{} &
\colhead{} &
\colhead{} &
\colhead{mJy} &
\colhead{mJy} &
\colhead{K} &
\colhead{$L_{\bigodot}$}    &
\colhead{}
}
\startdata
SMMJ00266$+$1708   & 18.6$\pm$2.4\tablenotemark{f} & 94$\pm$15\tablenotemark{f}
                 &22.5\tablenotemark{f}&I$>$26.1\tablenotemark{f}
                 &2.4&      & 2.73 &0.750&0.150&35.3& 0.5 &M82$+$AGN  \\
SMMJ02399$-$0136   & 23.0$\pm$1.9\tablenotemark{g} &526$\pm$10\tablenotemark{h}
                 &17.8\tablenotemark{g}&R=21.2\tablenotemark{g}
                 &2.5& 2.80 & 2.81 &1.12 &1.44&53.4& 2.5 &M82$+$AGN\\
SMMJ09429$+$4659   &  4.9$\pm$1.5\tablenotemark{i} &970$\pm$3.5\tablenotemark{j}
                 &19.7\tablenotemark{j}&R=25.2\tablenotemark{j}
                 &1.3&      & 2.38 &0.683&&83.0& 6.3 &M82       \\
SMMJ09431$+$4700   & 10.5$\pm$1.8\tablenotemark{i} & 55$\pm$3.5\tablenotemark{j}
                 &20.2\tablenotemark{j}&R=23.8\tablenotemark{j}
                 &1.2& 3.35 & 3.36 &0.607&1.04&41.6& 0.9 &M82$+$AGN\\
SMMJ10519$+$5723   & 4.5$\pm$1.3\tablenotemark{k}  & 47$\pm$10\tablenotemark{k}
                 &$>$20.4\tablenotemark{k}&I=24.6\tablenotemark{k}
                 &   & 2.69 & 2.67 &0.491&&42.0& 0.5 &M82 \\
SMMJ10521$+$5719   & 6.2$\pm$1.6\tablenotemark{k}  &278$\pm$12\tablenotemark{k}
                 &$>$20.6\tablenotemark{k}&I=22.7\tablenotemark{k}
                 &   & 2.69 & 2.69 &0.370&0.122&61.6& 2.9 &M82$+$AGN\\
MMJ120517$-$0743.1 & 6.3$\pm$0.9\tablenotemark{l}  &40$\pm$13\tablenotemark{m}
                 &22.5\tablenotemark{m}&R$\sim$25.4\tablenotemark{m}
                 &   &      & $>$3.6 &  & &  &  &   \\
MMJ120539$-$0745.4 & 6.3$\pm$1.4\tablenotemark{l}  &55$\pm$13\tablenotemark{m}
                 &$>$22.7\tablenotemark{m}&R$>$26.2\tablenotemark{m}
                 &   &      & $>$3.6 &  & &  &  &  \\
MMJ120546$-$0741.5 & 18.5$\pm$2.4\tablenotemark{l} &42$\pm$13\tablenotemark{m}
                 &21.9\tablenotemark{m}&R$>$26.2\tablenotemark{m}
                 &   &      & $>$3.6 &  & &  &  &  \\
MMJ154127$+$6615   & 10.7$\pm$1.2\tablenotemark{l} &81$\pm$13\tablenotemark{n}
                 &$>$21.2\tablenotemark{p}&R$>$24.9\tablenotemark{p}
                 &   &      & $>$3.6 &  & &  &  &  \\
MMJ154127$+$6616   & 14.6$\pm$1.8\tablenotemark{l} &67$\pm$13\tablenotemark{n}
                 &20.5\tablenotemark{n}&R$>$24.5\tablenotemark{n}
                 &   &      & 2.79 &0.472&0.159&35.0& 0.9 &M82 $+$AGN \\
SMMJ16369$+$4057   & 5.1$\pm$1.4\tablenotemark{k}  &74$\pm$29\tablenotemark{k}
                 &18.2\tablenotemark{k}&R=22.5\tablenotemark{k}
                 &   & 1.19 & 1.21 &0.493&0.205&27.5& 0.1 &M82$+$AGN\\
SMMJ16371$+$4053   &11.2$\pm2.9$\tablenotemark{o}  &74$\pm$23\tablenotemark{o}
                 &19.2\tablenotemark{q}&I=23.2\tablenotemark{q}
                 &   & 2.38 & 2.38 &0.519&0.242&34.3& 0.7 &M82$+$AGN\\
\enddata
\label{sample}

\begin{tabular} {l l l l l l l l l l l l}
$^\mathrm{a}$& \multicolumn{11}{l}{Adopted lensing magnification where applicable, magnification 1 is assumed otherwise. See text for references. The submm and radio fluxes and} \\
&\multicolumn{11}{l}{~optical/near-infrared magnitudes listed here are not corrected for amplification, while the infrared luminosities include the magnification correction.}\\
&\multicolumn{11}{l}{~Magnitudes are on the Vega system.}\\
$^\mathrm{b}$& \multicolumn{11}{l}{Redshift from previous measurements. See text for references.}\\
$^\mathrm{c}$& \multicolumn{11}{l}{Redshift from best fit to IRS mid-infrared spectrum (this work). Uncertainty 0.06 including effects of potential template mismatch (See Sect. 4} \\
&\multicolumn{11}{l}{~for details).}\\
$^\mathrm{d}$& \multicolumn{11}{l}{Flux of the PAH $7.7\,\mu{\rm m}$ rest frame feature after continuum subtraction.}\\
$^\mathrm{e}$&\multicolumn{11}{l}{Flux of the continuum $7.7\,\mu{\rm m}$ rest frame.}\\
$^\mathrm{f}$&\citet{frayer00}&
$^\mathrm{g}$&\citet{smail02}&
$^\mathrm{h}$&\citet{smail00}&
$^\mathrm{i}$&\citet{cowie02}& & & & \\
$^\mathrm{j}$&\multicolumn{10}{l}{\citet{ledlow02}. Opt/NIR magnitudes for SMMJ09431$+$4700 are dominated by the H6 component while H7 is the dominant submm component}\\
&\multicolumn{10}{l}{~\citep{tacconi06}.}\\
$^\mathrm{k}$&\citet{ivison02}&
$^\mathrm{l}$&\citet{eales03}&
$^\mathrm{m}$&\citet{dannerbauer04a}&
$^\mathrm{n}$&\citet{bertoldi00}&
$^\mathrm{o}$&\citet{chapman05}& &\\
$^\mathrm{p}$&\citet{dannerbauer04b}&
$^\mathrm{q}$&\citet{smail04}& & & & & & & &
\end{tabular}
\end{deluxetable}

\clearpage

\clearpage

\clearpage

\end{document}